\documentclass[twocolumn,english,pra]{revtex4-1}
\usepackage[T1]{fontenc}
\usepackage[latin9]{inputenc}
\usepackage{color}
\usepackage{float}
\usepackage{amsmath}
\usepackage{graphicx}
\usepackage{amssymb}
\usepackage{esint}

\makeatletter

\providecommand{\tabularnewline}{\\}
\newcommand{\lyxdot}{.}

\@ifundefined{textcolor}{}
{%
 \definecolor{BLACK}{gray}{0}
 \definecolor{WHITE}{gray}{1}
 \definecolor{RED}{rgb}{1,0,0}
 \definecolor{GREEN}{rgb}{0,1,0}
 \definecolor{BLUE}{rgb}{0,0,1}
 \definecolor{CYAN}{cmyk}{1,0,0,0}
 \definecolor{MAGENTA}{cmyk}{0,1,0,0}
 \definecolor{YELLOW}{cmyk}{0,0,1,0}
 }

\@ifundefined{definecolor}
 {\usepackage{color}}{}
\makeatother

\makeatother

\usepackage{babel}

\begin{document}

\title{Entanglement, EPR steering and Bell - nonlocality criteria for multipartite
higher spin systems }

\author{Q. Y. He, P. D. Drummond, and M. D. Reid}

\affiliation{ARC Centre of Excellence for Quantum-Atom Optics, Centre for Atom
Optics and Ultrafast Spectroscopy, Swinburne University of Technology,
Melbourne 3122, Australia}
\begin{abstract}
We develop criteria to detect three classes of nonlocality that have
been shown by Wiseman et al. {[}Phys. Rev. Lett.\textbf{ 98}, 140402
(2007){]} to be non-equivalent: entanglement, EPR steering, and the
failure of local hidden variable theories. We use the approach of
Cavalcanti et al. {[}Phys. Rev. Lett. \textbf{99}, 210405 (2007){]}
for continuous variables to develop the nonlocality criteria for arbitrary
spin observables defined on a discrete Hilbert space. The criteria
thus apply to multi-site qudits, i.e., systems of fixed dimension
$d$, and take the form of inequalities. We find that the spin moment
inequalities that test local hidden variables (Bell inequalities)
can be violated for arbitrary $d$ by optimised highly correlated
non-maximally entangled states provided the number of sites $N$ is
high enough. On the other hand, the spin inequalities for entanglement
are violated, and thus detect entanglement \textcolor{black}{for such
states,} for arbitrary $d$ and $N$, and with a violation that increases$\ $
with $N$. We show that one of the moment entanglement inequalities
can detect the entanglement of an arbitrary generalised multipartite
Greenberger-Horne-Zeilinger state. Because they involve the natural
observables for atomic systems, the relevant spin-operator correlations
should be readily observable in trapped ultra-cold atomic gases and
ion traps. 
\end{abstract}
\maketitle

\section{Introduction}

Entanglement and nonlocality has been a central issue in quantum mechanics
since the Einstein-Podolsky-Rosen (EPR) paradox \cite{epr} which
brought into focus the connection between entanglement and nonlocality.
The EPR paradox shows that there are correlated quantum states which
demonstrate an inconsistency between the completeness of quantum mechanics
and the concept of local realism. Schroedinger introduced the term
`steering' \cite{Schrodinger} to describe this apparent nonlocality,
and pointed out that these states necessarily involve entanglement,
that is, they cannot be separated into factorized terms. Later, Bell's
work \cite{Bell} served to demonstrate that the situation is even
more serious. Bell found quantum states whose correlations cannot
be explained by any possible local hidden variable (LHV) theory. Thus,
Einstein's hope of a local realistic completion of quantum mechanics
was not feasible.

Experimental demonstration of EPR and Bell inequality violations to
date has supported quantum mechanics. However, almost all work in
this direction to date has relied on rather small Hilbert spaces with
one or two massless particles. This leaves an important open question,
first raised by Schroedinger \cite{Schrodinger}. Will nonlocal quantum
phenomena still persist at a large scale? And do quantum superpositions
survive in this limit \cite{ystbeamsplit}? Here we note that there
are still many difficulties in unifying gravity with quantum mechanics,
and the large-scale existence of massive quantum entanglement would
directly test such unified theories \cite{Penrose}.

Ideally, one would like to generate quantum entanglement of distinct
mass distributions. This not only would test quantum theory in a new
domain, but also could lead to new types of sensitive gravity detectors.
To achieve this, a first step is to obtain macroscopic entanglement
of internal degrees of freedom in ultra-cold atoms, since low temperatures
are generally a prerequisite to the observation of quantum superpositions.
Already there has been much progress in this direction, with the generation
of entangled macroscopic samples of room-temperature atoms \cite{Plozik},
entanglement of ionic degrees of freedom \cite{Blatt}, and the observation
of spin-squeezing in an ultra-cold Bose-Einstein condensate \cite{Germany-spin&entanglement,atom-chip}.

While our motivation is to understand something more about Schroedinger's
cat, we focus in particular on the issue of macroscopic nonlocality.
This brings into question the usual idea of the classical-quantum
correspondence principle, that the system will revert to classical
local realism in the {}``large particle'' limit. For many particle
systems, macroscopic nonlocality remains to be explored in depth,
either experimentally or theoretically. In particular, it is essential
to understand the signatures of such effects, in terms of practically
accessible observable quantities. 

The three forms of nonlocality that we consider are \emph{entanglement}
\cite{ent}; \emph{steering} \cite{hw-1-1} (of which the EPR paradox
\cite{epr} is a special form; hence we follow Ref. \cite{ericsteer2}
in this paper and use the term {}``\emph{EPR steering}''); and failure
of local hidden variable (LHV) theories \cite{Bell}, which we call
\emph{Bell-nonlocality} \cite{ericsteer2}. For mixed states, these
forms of nonlocality are not equivalent \cite{hw-1-1}. The second
type of nonlocality, closely associated with the original EPR paradox
\cite{rrmp}, has received relatively little attention to date. Recent
work of Wiseman and co-workers \cite{hw-1-1,ericsteer2} formalises
steering as a nonequivalent nonlocality with its own experimental
criteria and has led to EPR steering being the subject of a recent
experimental investigation \cite{steerexp}.

On the theoretical side, much work has been done for Bell's nonlocality
on multi-site qubits \cite{mermin,ard,beklys}, and bipartite qudits
\cite{multibell,drumspinbell,peresspin,gsisspin,franmrspin,qudirincrezeil,qudit,fuqdit,Chen qudit,quditj},
though relatively little so far on multipartite systems of higher
dimensionality (qudits) \cite{Cabello,multisitequditson,multisitequdit}.
Our interest here is to explore these different types of {}``largeness,''
i.e., many sites, many dimensions, and the combination of both. It
is conventionally argued that a large system (either multiple sites
$N$ or higher dimensional $d$ at each site) must become consistent
with classical or LHV theories, but the extent to which this comes
about for the three forms of nonlocality is not clear. Whether classical
correspondence is achieved by simply increasing $d$, or $N$, or
whether it occurs through an increasing sensitivity to decoherence,
or some other mechanism, is a fundamental question.

Work on $N$-site qubits gave the surprising result that the deviation
of the quantum from the LHV theory increases exponentially with number
of qubits \cite{mermin}, for Greenberger-Horne-Zeilinger (GHZ) states
\cite{GHZ} and using the Mermin-Ardehali-Belinskii-Klyshko (MABK)
Bell inequalities \cite{mermin,ard,beklys}. The deviation is relatively
robust to noise and loss \cite{lossmermin}. However, with multiple
sites, it must be noted that the demonstration of entanglement or
nonlocality that is necessarily shared between all sites (true multipartite
nonlocality) is demonstrably more difficult \cite{multipartite}.
We will treat this type of entanglement in a subsequent publication.

One can generalise to qudits or higher-dimensional systems at each
site. Violations in the higher-dimensional case for bipartite systems
were found possible by Mermin \cite{multibell}, Drummond \cite{drumspinbell},
Peres and co-workers \cite{peresspin,gsisspin}, and Reid et al. \cite{franmrspin}.
More recent work has confirmed that the violation of LHV in the bipartite
case increases or is constant with dimensionality $d$ \cite{qudirincrezeil,qudit,fuqdit,Chen qudit,quditj}
and the CGLMP Bell inequalities (of Collins, Gisin, Linden, Massar,
and Popescu) \cite{qudit} have been shown to be tight. Multi-site
qudits have been examined by Cabello \cite{Cabello} and Son et al.
\cite{multisitequditson}, who extended the MABK inequalities, and
Chen et al. \cite{multisitequdit}, who developed tight inequalities
similar to the CGLMP inequalities for bipartite qudits. Results from
these authors indicate that the high-dimensional violations are steady,
or exponentially increasing, with the number of sites $N$.

We employ here the idea of studying the multi-site higher-dimensionality
nonlocality problem using moment inequalities of the type proposed
recently by Cavalcanti, Foster, Reid, and Drummond (CFRD) \cite{cvbell2}.
This originated in the work of Mermin, where the concept was mostly
restricted to multi-partite qubit measurements, and was applied to
Bell violations \cite{mermin}. The basis of the approach is to construct
complex operators $F_{k}^{\pm}=X_{k}\pm iP_{k}$ from two noncommuting
observables $X_{k}$ and $P_{k}$ defined for measurements at spatially
separated sites $k=1,...,N$. For any separable or local hidden variable
(LHV) model, the inequality\begin{equation}
|\langle\prod_{k=1}^{N}F_{k}^{\pm}\rangle|^{2}\leq\int_{\lambda}d\lambda P(\lambda)\prod_{k=1}^{N}|\langle F_{k}^{\pm}\rangle_{\lambda}|^{2}\label{eq:LHV}\end{equation}
then holds. Wiseman et al. \cite{hw-1-1} have developed separable
models in which \emph{some} of the sites are described by hidden variables
that are additionally required to be consistent with localised quantum
states. Cavalcanti et al. \cite{multipartitesteer} recently pointed
out that the nature of the separable model, whether we restrict to
local quantum or local hidden variable states at site $k$, enables
us to deduce a constraint on $|\langle F_{k}^{\pm}\rangle_{\lambda}|^{2}$,
and this leads to a criteria set involving the three types of nonlocality.
The case where local states at all sites are constrained to be quantum
states leads to the criteria for entanglement developed recently by
Hillery and co-workers \cite{hillzub,hilDN,spinhillery,mulithillery}. 

CFRD \cite{cvbell2} used Eq. (\ref{eq:LHV}) to develop a multi-site
continuous variable test of Bell nonlocality. They derived a Bell
inequality that is different from previous formulations because it
involves moments of the continuous variable outcomes and does not
assume bounded outcomes. Defining the outcomes of two \emph{arbitrary}
observables to be $X_{k}$ and $P_{k}$, at spatially separated sites
denoted by $k$ ($k=1,...,N$), these inequalities can be written\begin{eqnarray}
\Bigl|\langle\prod_{k=1}^{N}(X_{k}+iP_{k})\rangle\Bigr|^{2} & \leq & \langle\prod_{k=1}^{N}(X_{k}^{2}+P_{k}^{2})\rangle\,.\label{eq:bellcfrd}\end{eqnarray}
 The left side ($L$) of the inequality is measured by way of the
moments involving the Hermitian observables defined at each site:
e.g. for $N=2$, $L=\langle X_{1}X_{2}-P_{1}P_{2}\rangle^{2}+\langle P_{1}X_{2}+X_{1}P_{2}\rangle^{2}$.
Multi-observable and higher order extensions of these inequalities
have been presented by Shchukin and Vogel \cite{vogel} and Sun et
al. \cite{zu}. CFRD showed that for {}``position'' and {}``momentum''
measurements with a continuous spectrum, there is a violation of Eq.
(\ref{eq:bellcfrd}) predicted for multipartite qubit GHZ states \cite{GHZ}.
Recent work extended the CFRD to examine inequalities involving optimisation
of functions of these observables that give less strict requirements
for loss \cite{function cfrd}. The main point is that because $X_{k}$
and $P_{k}$ are arbitrary observables, a ready generalisation exists
to higher-spin observables.

In this paper, we thus develop the CFRD approach further by using
arbitrary spin operators, which are generic observables that can be
measured in many physical environments for systems with discrete states.
We also apply the unified method of Cavalcanti et al. \cite{multipartitesteer}
to derive CFRD-type criteria for multipartite entanglement, EPR steering,
and Bell nonlocality for $N$-site systems of higher dimensionality
$d$ (i.e. specifically, for higher spin $J$). Cavalcanti et al.
\cite{multipartitesteer} previously examined only the multipartite
qubit ($J=1/2$) case, but used a method similar to the original approach
of Mermin \cite{mermin} that allows stronger criteria to be derived
in this scenario. ~Direct application of the Bell inequality (\ref{eq:bellcfrd})
without optimisation shows demonstrations of Bell nonlocality possible
for maximally entangled states where \textcolor{black}{$d=2,\ 3$,
a}nd for all $N\geq3$. Certain non-maximally entangled but highly
correlated states of the type considered by Acin et al. \cite{acinthreelevel}
allow violations for all $d$ provided the number of sites is large
enough (e.g., $d=5$, requires $N\geq9$). While the demonstration
of Bell nonlocality is limited in terms of dimensionality $d$ for
low $N$ (whereas the CGLMP-type Bell inequalities give constant or
increasing violation for bipartite states), the violations that we
find for optimised non-maximally entangled states increase with respect
to the number of sites $N$. 

To derive entanglement and EPR steering criteria, we use the lower
bound $C_{J}$ of the quantum uncertainty relation\begin{equation}
\Delta^{2}\hat{J}^{x}+\Delta^{2}\hat{J}^{y}\geq C_{J}\ \label{eq:spin1/2ur-1-1}\end{equation}
for two conjugate spins $\hat{J}^{x}$ and $\hat{J}^{y}$ \cite{hoftoth}
with fixed total spin $J$ that has been derived recently for all
\textcolor{black}{$J$ ($d$) \cite{cj}. For optimised non-maximally
entangled states, the violation of the {}``$C_{J}$-entanglement
inequality'' occurs for all spin (dimensionality) $J$ and $N$,
and decreases with $J$ but increases with $N$. A similar result
was obtained for GHZ states by Roy for the spin $J=1/2$ case \cite{royprl}.
We also derive a second set of more general entanglement and EPR steering
criteria based on the original Heisenberg uncertainty relation. This
set therefore does not assume the case of fixed total spin $J$, and
for entanglement is a generalisation of criteria developed recently
by Hillery and co-workers \cite{hillzub,hilDN,mulithillery,spinhillery}.
In this paper, we discuss the use of the two types of criteria for
both maximally entangled and non-maximally entangled highly correlated
spin states.}

\section{The local hidden variable (LHV) and Local Hidden State (LHS) Models}

We consider measurements $\hat{X}_{k}$, with associated outcomes
$X_{k}$, that can be performed on the $k$-th system ($k=1,...,N$).
Following Bell \cite{Bell}, we have a \emph{local hidden variable
model} (LHV) if the joint probability for outcomes of simultaneous
measurements performed on the $N$ spatially separated systems is
given by\begin{equation}
P(X_{1},...,X_{N})=\int_{\lambda}P(\lambda)P(X_{1}|\lambda)...P(X_{N}|\lambda)d\lambda\,.\label{eq:bell}\end{equation}
Here $\lambda$ are Bell's local hidden variables, and $P(X_{k}|\lambda)$
is the probability of $X_{k}$ given the values of $\lambda$, with
$P(\lambda)$ being the probability distribution for $\lambda$. The
factorisation in the integrand is Bell's locality assumption, that
$P(X_{k}|\lambda)$ depends on the parameters $\lambda$, and the
measurement choice made at $k$ only. The hidden variables $\lambda$
describe a\emph{ }local state\emph{ }for each site, in that the probability
distribution $P(X_{k}|\lambda)$ for the measurement at $k$ is given
as a function of the $\lambda$. If (\ref{eq:bell}) fails, then we
have\emph{ }proved\emph{ }a\emph{ }failure of LHV theories\emph{,
}which we generically term a \emph{Bell violation }or\emph{ Bell nonlocality}
\cite{ericsteer2}\emph{. }

Bell's locality does not exclude that the local hidden state could
be a quantum state, in which case there exists a quantum density operator
$\rho_{k}$ for which $P(X_{k}|\lambda)=\langle X_{k}|\rho_{k}|X_{k}\rangle$.
In this case, we write $P(X_{k}|\lambda)\equiv P_{Q}(X_{k}|\lambda)$,
where the subscript $Q$ denotes the quantum probability distribution.
When all the $P(X_{k}|\lambda)$ ($k=1,...,N$) are so restricted,
we write\begin{equation}
P(X_{1},...,X_{N})=\int_{\lambda}P(\lambda)P_{Q}(X_{1}|\lambda)...P_{Q}(X_{N}|\lambda)d\lambda\,,\label{eq:sepent}\end{equation}
which is the requirement of a \emph{quantum separable }(QS)\emph{
model} \cite{hw-1-1,ericsteer2}. The model (\ref{eq:sepent}) follows
from the assumption of a separable density operator $\rho$, which
can be written in the factorised form $\rho=\sum_{R}P_{R}\rho_{1}^{R}...\rho_{N}^{R}$,
\textcolor{black}{where $R$ describes a set of local quantum states
for each site.} Failure of Eq. (\ref{eq:sepent}) gives proof of entanglement\emph{,}
following standard definitions \cite{ent}.

It is clear that all QS models are therefore a subset of the LHV class,
that is, quantum separable models are a special case of general local
hidden variable theories, so that one can write\begin{equation}
\left\{ QS\right\} \implies\left\{ LHV\right\} \,.\end{equation}
This means in turn that failure of $\left\{ LHV\right\} $ also implies
failure of $\left\{ QS\right\} $, which is termed entanglement. Another
way to state this is that entanglement is a necessary condition for
a Bell violation.

Wiseman et al. \cite{hw-1-1} have pointed out that there exists an
intermediate case between the local hidden variable (LHV) and quantum
separable (QS) models, in which for the bipartite case $N=2$ one
of the $P(X_{k}|\lambda$) is constrained to be a quantum distribution
and the other is not. Failure of this asymmetric Local Hidden State
(LHS) model was shown by them to demonstrate Schroedinger's {}``steering''.
The connection between this model and Schroedinger's {}``steering''
and the EPR paradox for $N=2$ has been explained in Ref. \cite{hw-1-1}
and Ref. \cite{ericsteer2} and is summarised in terms of an {}``elements
of reality'' approach in Ref. \cite{rrmp}. Where $N=2$ and $T=1$,
we arrive at a model which if violated is a demonstration of {}``steering'',
and also is a demonstration of the EPR paradox as generalised to appropriate
observables \cite{rrmp}; hence we follow Ref. \cite{ericsteer2}
and use in this paper the term {}``EPR-steering'' to describe failure
of this model. 

Recent work of Cavalcanti et al. \cite{multipartitesteer} generalises
the LHS model to multiple sites. Following them, when exactly $T$
of the $P(X_{k}|\lambda)$ ($k=1,...,N$) of the separable model (\ref{eq:bell})
are quantum probabilities, one can write (we label these $T$ sites
by $k=1,...,N$)\begin{equation}
P(X_{1},...,X_{N})=\int_{\lambda}P(\lambda)\prod_{k=1}^{T}P_{Q}(X_{T}|\lambda)...\prod_{k=T+1}^{N}P(X_{N}|\lambda)d\lambda\,.\label{eq:sepent-1}\end{equation}
This condition implies that we assume normal quantum uncertainties
for $X_{1},...,X_{T}$, while for the remaining observables we can
have a complete classical knowledge, i.e., they are predetermined
elements of local reality in Einstein's language. In this paper, we
refer to the multipartite separability model (\ref{eq:sepent-1})
as a $T$-th order EPR model (EPR$_{T}$) and follow \cite{multipartitesteer}
to denote this Local Hidden State model by LHS(T,N). With $T=N$ one
has a simple case of proving entanglement, while with $T=0$ one has
a Bell violation. Importantly for this paper, Cavalcanti et al. \cite{multipartitesteer}
show that the case of violation of LHS(1,N) ($T=1$) implies EPR-steering
nonlocality across at least one bipartition.

\begin{figure}[H]
\begin{centering}
\includegraphics[width=0.6\columnwidth]{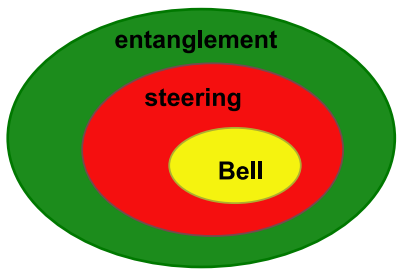} 
\par\end{centering}

\caption{\label{fig:Venn}Three famous types of {}``quantum nonlocality\textquotedblright{}:
Bell nonlocality is a stronger result than {}``EPR steering'', which
is stronger than entanglement. }

\end{figure}

For larger $N$, we have a range of separable models as has been shown
by Ref. \cite{multipartitesteer}: violations of these models provide
a step-by-step transition in increasing degrees of \emph{quantum nonlocality}.
It is clear that all $\left\{ EPR_{T}\right\} $ (i.e. LHS(T,N)) models
are included in the LHV class, so that one can write\begin{equation}
\left\{ QSM\right\} \implies\left\{ EPR_{T+1}\right\} \implies\left\{ EPR_{T}\right\} \implies\left\{ LHV\right\} \,,\end{equation}
 and hence when violations are observed, one similarly obtains the
negation of these relations\begin{equation}
\left\{ Bell\right\} \implies\left\{ S_{T}\right\} \implies\left\{ S_{T+1}\right\} \implies\left\{ entanglement\right\} \,,\end{equation}
where $\{S_{T}\}$ symbolises the nonlocality associated with the
failure of the $\{EPR_{T}\}$ (i.e. LHS(T,N)) model. This is shown
in a Venn diagram in Fig. \ref{fig:Venn} for the $N=2$ or bipartite
case. More generally there is a nested sequence of nonlocality inequalities.
For the bipartite case, Werner \cite{werner} has proved that the
entangled states are a strict superset of Bell states (those that
show Bell nonlocality). Wiseman et al. \cite{hw-1-1} proved that
those states able to show steering are a strict superset of Bell states,
and a strict subset of entangled states, and hence showed that there
are three distinct classes as illustrated in the Venn diagram in Fig.
\ref{fig:Venn}. 

\section{\textbf{Derivation of a Spin nonlocality criteria set}}

We are considering $N$ sites that are in principle causally separated.
From previous references summarised in the last section, we know that
depending on the assumption of what type of local hidden state is
present at each site, whether {}``quantum'' or {}``hidden variable'',
one can derive criteria for Bell-nonlocality, EPR-steering, or entanglement
\cite{ericsteer2,multipartitesteer}. There are many possible observable
signatures for these types of nonlocality. Generally speaking it is
simplest to construct conditions \emph{sufficient} to deduce the nonlocality,
by deriving inequalities for observed correlations that follow \emph{necessarily
}from the LHV, LHS(1,N) and QS models, respectively. This is the route
we take here. Our criteria will thus take the form of inequalities
that if violated prove the nonlocality, but will not necessarily be
violated by all such nonlocal states.

We will use the general results of Cavalcanti et al. \cite{multipartitesteer}
to derive nonlocality inequalities for higher-spin measurements. We
follow their derivation, using their notation as much as possible,
but applying to our special case of arbitrary spin observables. First,
one follows Mermin \cite{mermin} and Cavalcanti et al. \cite{cvbell2}
who consider $F_{k}^{\pm}=(X_{k}\pm iP_{k})$ for each site $k$,
where $X_{k}$ and $P_{k}$ denote the outcomes of two observables
$\hat{X_{k}}$ and $\hat{P_{k}}$. Now we turn to the specific case
of interest in this paper, and consider spin measurements, at each
site. Here, we make the simplest choice:\begin{equation}
F_{k}^{\pm}=J_{k}^{\theta}\pm iJ_{k}^{\theta'}\,,\end{equation}
where $J_{k}^{\theta}=J_{k}^{x}\cos\theta+J_{k}^{y}\sin\theta$, and
$J_{k}^{x/y}$ are the outcomes for the measurements represented by
spin operators $\hat{J_{k}^{x}}$ and $\hat{J_{k}^{y}}$. In fact
we will focus on the case where $\theta'=\theta+\pi/2$ for which
$F_{k}^{\pm}=J^{\pm}$ corresponds to the spin raising and lowering
operators $\hat{J}^{\pm}$, i.e., \begin{equation}
F_{k}^{\pm}\equiv J_{k}^{\pm}=J_{k}^{x}\pm iJ_{k}^{y}\,,\end{equation}
which is the choice $X_{k}=J_{k}^{x}$ and $P_{k}=J_{k}^{y}$. For
notational convenience, we thus denote the raising operator $\hat{J^{\dagger}}$
by $\hat{J}^{+}$ and the lowering operator $\hat{J}$ by $\hat{J}^{-}$,
and their outcomes by $J^{+}$ and $J^{-}$, respectively. Note the
distinction between the operator $\hat{J_{k}}^{\pm}$ and the \emph{measurable
complex number} $\langle J_{k}^{\pm}\rangle=\langle J_{k}^{x}\rangle\pm i\langle J_{k}^{y}\rangle$.

Following Ref. \cite{multipartitesteer}, for any LHV or LHS model
(\ref{eq:sepent-1}), we can write \begin{equation}
\langle\prod_{k=1}^{N}J_{k}^{s_{k}}\rangle=\int_{\lambda}d\lambda P(\lambda)\prod_{k=1}^{N}\langle J_{k}^{s_{k}}\rangle_{\lambda}\,,\label{eqn:sepave}\end{equation}
where $s_{k}=+1$ or $-1$. Here $\langle J_{k}^{\pm}\rangle_{\lambda}=\langle J_{k}^{x}\rangle_{\lambda}\pm i\langle J_{k}^{y}\rangle_{\lambda}$
where the subscript $\lambda$ denotes the complex number average,
for a given hidden variable specification $\lambda$. Then one follows
the Holder inequality techniques of Refs. \cite{cvbell2,hillzub}
and uses the inequality (\ref{eq:LHV}), which holds since\begin{eqnarray}
|\langle\prod_{k=1}^{N}F_{k}^{\pm}\rangle| & \leq & \int d\lambda P(\lambda)|\langle F_{1}^{\pm}\rangle_{\lambda}||\langle F_{2}^{\pm}\rangle_{\lambda}|...\nonumber \\
 & = & \int d\lambda P(\lambda)\Bigl[|\langle F_{1}^{\pm}\rangle_{\lambda}|^{2}|\langle F_{2}^{\pm}\rangle_{\lambda}|^{2}...\Bigr]^{1/2}\nonumber \\
 & \leq & [\int d\lambda P(\lambda)]^{1/2}\nonumber \\
 &  & \,\,\times[\int d\lambda P(\lambda)|\langle F_{1}^{\pm}\rangle_{\lambda}|^{2}|\langle F_{2}^{\pm}\rangle_{\lambda}|^{2}...]^{1/2}\nonumber \\
 & = & [\int d\lambda P(\lambda)|\langle F_{1}^{\pm}\rangle_{\lambda}|^{2}|\langle F_{2}^{\pm}\rangle_{\lambda}|^{2}...]^{1/2}\,.\label{eq:deriv}\end{eqnarray}
Here, the Cauchy-Schwarz inequality $\langle xy\rangle^{2}\leq\langle x^{2}\rangle\langle y^{2}\rangle$
where $x=\sqrt{P(\lambda)}$ and $y=\sqrt{P(\lambda)}|\langle F_{1}^{\pm}\rangle_{\lambda}||\langle F_{2}^{\pm}\rangle_{\lambda}|...$
has been used to justify the third line.

\begin{figure}[H]
\begin{centering}
\includegraphics[clip,width=0.5\columnwidth]{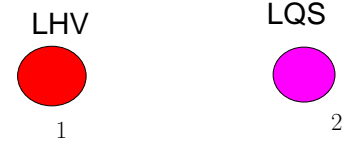}
\par\end{centering}

\caption{\label{fig:Venn-1-2}The hybrid model of Ref. \cite{hw-1-1} involves
different {}``local hidden states'', either quantum (LQS) or local
hidden variable (LHV), at each spatially separated site $1$ and $2$.
The asymmetric use of quantum uncertainty relations that results because
of this gives rise to criteria for steering and the EPR paradox ({}``EPR
steering'') \cite{ericsteer2}.}

\end{figure}

By definition $|\langle J_{k}^{\pm}\rangle_{\lambda}|^{2}=\langle J_{k}^{x}\rangle_{\lambda}^{2}+\langle J_{k}^{y}\rangle_{\lambda}^{2}$,
since variances are non-negative it follows that for any \emph{local
hidden variable theory} $\lambda$,\begin{equation}
\langle(J_{k}^{x})^{2}\rangle_{\lambda}-\langle J_{k}^{x}\rangle_{\lambda}^{2}\geq0,\,\langle(J_{k}^{y})^{2}\rangle_{\lambda}-\langle J_{k}^{y}\rangle_{\lambda}^{2}\geq0\,,\label{eq:positive variance}\end{equation}
and hence that\begin{equation}
|\langle J_{k}^{\pm}\rangle_{\lambda}|^{2}\leq\langle(J_{k}^{x})^{2}\rangle_{\lambda}+\langle(J_{k}^{y})^{2}\rangle_{\lambda}\,.\label{eq:lhv1}\end{equation}
The case where the separable model specifies $N$ local \emph{quantum}
states, as in the assumption of the separable density operator (\ref{eq:sepent}),
has been employed for example by Roy \cite{royprl}. Here there are
further restrictions due to the Heisenberg uncertainty principle and
its generalizations. For spins, $ $$\Delta J_{k}^{x}\Delta J_{k}^{y}\geq|\langle J_{k}^{z}\rangle|/2$
and hence\begin{equation}
(\Delta J_{k}^{x})^{2}+(\Delta J_{k}^{y})^{2}\geq|\langle J_{k}^{z}\rangle|\,.\label{eq:qurcom}\end{equation}
Quantum uncertainty relations of the form \begin{equation}
(\Delta J_{k}^{x})^{2}+(\Delta J_{k}^{y})^{2}\geq C_{k}\,,\label{eq:lower bound}\end{equation}
where $C_{k}$ is a constant, can also be derived that give useful
entanglement and steering criteria, as will be introduced in the following
parts. This leads to the inequalities\begin{eqnarray}
|\langle J_{k}^{\pm}\rangle_{\lambda}|^{2} & \leq & \langle(J_{k}^{x})^{2}\rangle_{\lambda}+\langle(J_{k}^{y})^{2}\rangle_{\lambda}-C_{k}\,,\label{eq:lqs3}\end{eqnarray}
\begin{eqnarray}
|\langle J_{k}^{\pm}\rangle_{\lambda}|^{2} & \leq & \langle(J_{k}^{x})^{2}\rangle_{\lambda}+\langle(J_{k}^{y})^{2}\rangle_{\lambda}-|\langle J_{k}^{z}\rangle_{\lambda}|\,.\label{eq:lqs3-1}\end{eqnarray}
The last inequality in fact implies \begin{eqnarray}
|\langle J_{k}^{+}\rangle_{\lambda}|^{2} & \leq & \langle(J_{k}^{x})^{2}\rangle_{\lambda}+\langle(J_{k}^{y})^{2}\rangle_{\lambda}\pm\langle J_{k}^{z}\rangle_{\lambda}\,,\label{eq:lqs3-1-1}\end{eqnarray}
\begin{eqnarray}
|\langle J_{k}^{-}\rangle_{\lambda}|^{2} & \leq & \langle(J_{k}^{x})^{2}\rangle_{\lambda}+\langle(J_{k}^{y})^{2}\rangle_{\lambda}\pm\langle J_{k}^{z}\rangle_{\lambda}\,.\label{eq:lqs3-1-2}\end{eqnarray}
We now use the results of Ref. \cite{multipartitesteer}: we assume
the model LHS(T,N) where sites $k=1,...,T$ are quantum, and the remainder
local hidden variable, so we have a hybrid case as studied for $N=2$
in Ref. \cite{hw-1-1} (Fig. \ref{fig:Venn-1-2}). Using the relations
then the following holds: \begin{multline}
|\langle\prod_{k=1}^{N}J_{k}^{s_{k}}\rangle|^{2}\leq\int d\lambda P(\lambda)\prod_{k=1}^{N}|\langle J_{k}^{s_{k}}\rangle_{\lambda}|^{2}\\
\leq\left\langle \prod_{k=1}^{T}(J_{k}^{x})^{2}+(J_{k}^{y})^{2}-C_{k})\prod_{k=T+1}^{N}(J_{k}^{x})^{2}+(J_{k}^{y})^{2}\right\rangle \,.\label{eq:ineqcom}\end{multline}
Here $s_{k}$ can be selected $+$ or $-$ at both sides, respectively.
If $T=0$ one recovers a Bell inequality whose violation will prove
failure of LHV, while for $T=N$, one recovers an inequality which
if violated will simply prove entanglement. The intermediate case
of Ref. \cite{multipartitesteer}, where $T=1$, recovers an inequality
which if violated will prove an EPR-steering. It is clearly necessary
to have entanglement as a starting point toward observation of stronger
forms of nonlocality. Similarly, using Eqs. (\ref{eq:lqs3-1-1}),
(\ref{eq:lqs3-1-2}), and (\ref{eq:lhv1}), the result of Ref. \cite{multipartitesteer}
becomes\begin{multline}
|\langle\prod_{k=1}^{N}J_{k}^{s_{k}}\rangle|^{2}\leq\int d\lambda P(\lambda)\prod_{k=1}^{N}|\langle J_{k}^{s_{k}}\rangle_{\lambda}|^{2}\\
\leq\left\langle \prod_{k=1}^{T}(J_{k}^{x})^{2}+(J_{k}^{y})^{2}\pm J_{k}^{z})\prod_{k=T+1}^{N}(J_{k}^{x})^{2}+(J_{k}^{y})^{2}\right\rangle \,,\label{eq:ineqcom-1}\end{multline}
where the $\pm$ appearing in the first $T$ factors of the right-side
product can be chosen independently for each factor. 

\section{\textbf{spin nonlocality criteria}}

In general, the total spin may itself be an observable, so that $J$
at each site is not known in advance. In this general case, we note
for all quantum states, we must have $|\langle\hat{J}^{+}\rangle||\langle\hat{J}^{-}\rangle|\leq\langle\hat{J}^{+}\hat{J}^{-}\rangle$
and $|\langle\hat{J}^{+}\rangle||\langle\hat{J}^{-}\rangle|\leq\langle\hat{J}^{-}\hat{J}^{+}\rangle$.
This implies\begin{equation}
|\langle\hat{J}^{+}\rangle|^{2}\leq\langle\hat{J}^{\pm}\hat{J}^{\mp}\rangle=\langle(\hat{J}^{x})^{2}+(\hat{J}^{y})^{2}\pm i[\hat{J}^{x},\hat{J}^{y}]\rangle\,,\end{equation}
which is another way to arrive at the conditions \eqref{eq:lqs3-1-1}-\eqref{eq:lqs3-1-2}.
Using \eqref{eq:ineqcom-1} we now obtain three nonlocality inequalities
that apply to all systems, with no assumptions being placed on the
total spin.

\subsection{Entanglement Inequalities: the generalised HZ entanglement criterion}

Entanglement is verified if\begin{eqnarray}
|\langle\prod_{k=1}^{N}\hat{J}_{k}^{s_{k}}\rangle|^{2} & > & \langle\prod_{k=1}^{N}[(\hat{J}_{k})^{2}-(\hat{J}_{k}^{z})^{2}\pm\hat{J}_{k}^{z}]\rangle\nonumber \\
 & \equiv & \langle\prod_{k=1}^{N}[(\hat{J}_{k})^{2}-(\hat{J}_{k}^{z})^{2}+l_{k}\hat{J}_{k}^{z}]\nonumber \\
 & = & \langle\prod_{k=1}^{N}\hat{J}_{k}^{\pm}\hat{J}_{k}^{\mp}\rangle\equiv\langle\prod_{k=1}^{N}\hat{J}_{k}^{l_{k}}\hat{J}_{k}^{-l_{k}}\rangle\,,\label{eq:entcrit}\end{eqnarray}
where $l_{k,}s_{k}=\pm$ and we note that $(\hat{J}_{1})^{2}-(\hat{J}_{1}^{z})^{2}\pm\hat{J}_{1}^{z}=\hat{J}_{1}^{\pm}\hat{J}_{1}^{\mp}$,
so the final line has been rewritten in terms of the lowering and
raising operators. The $\pm$ value of $l_{k}$ in each factor on
the right side ($R$) can be selected \emph{independently} for each
factor, and independently of the choice $s_{k}$, in order to minimise
the $R$. Also, the choice of $+$ or $-$ for $s_{k}$ on the left
side ($L$) can be selected independently to optimize the criterion
for the state used. 

We call this a \emph{generalized} HZ criterion, since a similar but
not identical criterion (\ref{eq:entcrit}) has been derived recently
by Hillery and co-workers \cite{hillzub,hilDN,spinhillery,mulithillery}.
This HZ multipartite criterion is an extension of the criteria developed
previously by Hillery and Zubairy \cite{hillzub}, and Hillery, Dung
and Niset \cite{hilDN}. There have been recent applications of this
criterion to spins systems \cite{mulithillery,spinhillery}. We recall
that $\hat{J}^{2}$ is defined as\begin{equation}
(\hat{J})^{2}=(\hat{J}^{x})^{2}+(\hat{J}^{y})^{2}+(\hat{J}^{z})^{2}\,,\end{equation}
so we can use $(\hat{J})^{2}-(\hat{J}^{z})^{2}=(\hat{J}^{x})^{2}+(\hat{J}^{y})^{2}$
to re-express the generalised HZ entanglement criterion (\ref{eq:entcrit})
as\begin{eqnarray}
|\langle\prod_{k=1}^{N}\hat{J}_{k}^{s_{k}}\rangle|^{2} & > & \langle\prod_{k=1}^{N}[(\hat{J}_{k}^{x})^{2}+(\hat{J}_{k}^{y})^{2}\pm\hat{J}_{k}^{z}]\,.\label{eq:entxy}\end{eqnarray}

\subsection{EPR-steering inequalities}

An EPR paradox or steering nonlocality is demonstrated if the EPR
steering inequality is violated, i.e., if \begin{equation}
|\langle\prod_{k=1}^{N}J_{k}^{s_{k}}\rangle|^{2}>\langle[(J_{1}^{2})^{2}-(J_{1}^{z})^{2}\pm J_{1}^{z}]\prod_{k=2}^{N}[(J_{k}^{x})^{2}+(J_{k}^{y})^{2}]\rangle\,.\label{eq:sterrcrit}\end{equation}
We note in this case $T=1$ so there is only one {}``quantum'' site
(which we select to be $k=1$). The criterion can also be expressed
as\begin{equation}
|\langle\prod_{k=1}^{N}J_{k}^{s_{k}}\rangle|^{2}>\langle[(J_{1}^{x})^{2}+(J_{1}^{y})^{2}\pm J_{1}^{z}]\prod_{k=2}^{N}[(J_{k}^{x})^{2}+(J_{k}^{y})^{2}]\rangle\label{eq:sterrcrit-1}\end{equation}
for direct comparison with (\ref{eq:entxy}). The criterion to detect
failure of the LHS(T,N) model (\ref{eq:sepent-1}) is \begin{equation}
|\langle\prod_{k=1}^{N}J_{k}^{s_{k}}\rangle|^{2}>\langle\prod_{k=1}^{T}[(J_{k}^{x})^{2}+(J_{k}^{y})^{2}\pm J_{k}^{z}]\prod_{k=T+1}^{N}[(J_{k}^{x})^{2}+(J_{k}^{y})^{2}]\rangle\,.\label{eq:sterrcrit-1-1}\end{equation}

\subsection{Bell-nonlocality inequalities}

A Bell inequality is violated, proving a failure of all possible LHV,
if:\begin{equation}
|\langle\prod_{k=1}^{N}J_{k}^{s_{k}}\rangle|^{2}>\langle\prod_{k=1}^{N}[(J_{k}^{x})^{2}+(J_{k}^{y})^{2}]\rangle\,.\label{eq:bellcrit}\end{equation}
We note the criterion has been expressed in terms of the moments of
outcomes corresponding to the observables. This is done because the
test of nonlocality is a test of local hidden variable theories and
hence does not assume quantum mechanics. The EPR-steering criterion
is expressed similarly in terms of outcome moments, for similar reasons.

This Bell inequality differs from the MABK Bell inequalities \cite{mermin,ard,beklys}
because the right side is not a fixed bound, but varies as a moment.
We will note that for spin-$1/2$ observables this Bell inequality
does not give as strong a violation as the MABK Bell inequalities
for maximally entangled states. The different nature of the right
hand side may however make the violations stronger in other scenarios.

These results (\ref{eq:entcrit})-(\ref{eq:bellcrit}) apply for all
spin measurements and systems, even when the spin quantum number itself
is a quantum observable.

\section{Fixed-dimensionality $J$ entanglement and EPR-Steering criteria}

We now consider states of fixed spin dimensionality $J$. The most
general pure quantum state of this type at a single site is simply
a general qudit state of dimension $d=2J+1$:\begin{eqnarray}
|\psi\rangle & = & \frac{1}{\sqrt{n}}[r_{-J}e^{-i\phi_{-J}}|J,-J\rangle+r_{-J+1}e^{-i\phi_{-J+1}}|J,-J+1\rangle\nonumber \\
 &  & \ \ \ \ +...+r_{J}e^{-i\phi_{J}}|J,+J\rangle]\,.\label{eq:purestate}\end{eqnarray}
\textcolor{black}{Here $n={\displaystyle \sum_{m=-J}^{J}}r_{m}^{2}$,
$r_{m},\phi_{m}\ (m=-J,...,J)$ are real numbers indicating amplitude
and phase respectively. }Where we have fixed dimensionality, i.e.
a spin-$J$ system, the quantum uncertainty relation (\ref{eq:lower bound})
can be used to derive different entanglement and steering inequalities
beyond those of (\ref{eq:entcrit})-(\ref{eq:sterrcrit-1}). In particular
there will be a quantum uncertainty relation of the \textcolor{black}{form
\cite{cj}}\begin{equation}
\Delta^{2}\hat{J}^{x}+\Delta^{2}\hat{J}^{y}\geq C_{J}\,,\label{eq:spin1/2ur-1}\end{equation}
where $C_{J}\neq0$ because there exist no simultaneous eigenstates
of $\hat{J}^{x}$ and $\hat{J}^{y}$ \cite{hoftoth}. The results
for selected values of $J$ are tabulated in Table (\ref{tab:Lower-bound-cj})
\cite{cj}.

\begin{table}[H]
\begin{centering}
\begin{tabular}{|c|c|c|c|c|c|c|c|c|c}
\hline 
$J$  & $1/2$ & $1$  & $3/2$  & $2$  & $5/2$  & $3$  & $7/2$  & $4$  & $\ldots$\tabularnewline
\hline
\hline 
$C_{J}$  & $1/4$  & $7/16$  & $0.6009$  & $0.7496$  & $0.8877$  & $1.0178$  & $1.1416$  & $1.26$  & $\ldots$\tabularnewline
\hline
\end{tabular}
\par\end{centering}

\caption{Lower bound $C_{J}$ of quantum uncertainty relation with spin-$J$,
where $C_{J}=7/16$ for spin-$1$ agrees with the result in \cite{hoftoth}.
\label{tab:Lower-bound-cj}}

\end{table}

These fixed-$J$ uncertainty relations can be used to derive additional
entanglement and EPR-steering criteria based on Eq. (\ref{eq:ineqcom}):
\begin{enumerate}
\item Entanglement is verified if\begin{eqnarray}
|\langle\prod_{k=1}^{N}J_{k}^{s_{k}}\rangle|^{2} & > & \langle\prod_{k=1}^{N}[(J_{k})^{2}-(J_{k}^{z})^{2}-C_{J}]\rangle\,.\label{eq:spinjent}\end{eqnarray}

\item An EPR-steering nonlocality is verified if \begin{eqnarray}
|\langle\prod_{k=1}^{N}J_{k}^{s_{k}}\rangle|^{2} & > & \langle[(J_{1})^{2}-(J_{1}^{z})^{2}-C_{J}]\nonumber \\
 &  & \ \ \times\prod_{k=2}^{N}[(J_{k}^{x})^{2}+(J_{k}^{y})^{2}]\rangle\,,\label{eq:spinjsteer}\end{eqnarray}

\end{enumerate}
where again we note the that $(\hat{J})^{2}-(\hat{J}^{z})^{2}=(\hat{J}^{x})^{2}+(\hat{J}^{y})^{2}$.
The criterion to detect failure of the more general LHS(T,N) separable
model (\ref{eq:sepent-1}) of \cite{multipartitesteer} is:\begin{eqnarray}
|\langle\prod_{k=1}^{N}J_{k}^{s_{k}}\rangle|^{2} & > & \langle\prod_{k=1}^{T}[(J_{k})^{2}-(J_{k}^{z})^{2}-C_{J}]\nonumber \\
 &  & \ \ \times\prod_{k=T+1}^{N}[(J_{k}^{x})^{2}+(J_{k}^{y})^{2}]\rangle\,.\label{eq:spinjsteer-1}\end{eqnarray}
We will refer to these criteria throughout the paper as the {}``$C_{J}"$
nonlocality criteria. 

The Bell inequality (\ref{eq:bellcrit}), which is not dependent on
$C_{J}$, also applies. The criteria we have derived here are all
\emph{sufficient} to detect entanglement, but may not \emph{necessarily
}detect entanglement. In the following, we analyze these spin nonlocality
sets in greater detail, and examine how the sensitivity to entanglement
changes as $J$ changes. We note here that either form of entanglement
and EPR-steering nonlocality criterion is valid for systems of fixed
dimensionality (i.e., $J$), and the optimal choice that is made will
depend on the value of $J$ and the states that are selected.

\section{Fixed-$J$ entangled states}

\textcolor{black}{There are many possible entangled states. In particular,
in the following sections we choose to analyze the following highly
correlated spin states:} 
\begin{enumerate}
\item \textcolor{black}{The maximally entangled and highly correlated states
of form\begin{multline}
|\Psi\rangle_{max}=\frac{1}{\sqrt{d}}\sum_{m=-J}^{J}|J,m\rangle_{1}|J,m\rangle_{2}|J,m\rangle_{3}...\,,\label{eq:maximally entangled state}\end{multline}
where $|J,m\rangle$ is an eigenstate of $\hat{J}^{2}$ and $\hat{J}^{z}$.
In particular the bipartite ($N=2$) maximally entangled state is\begin{equation}
|\Psi\rangle_{max}=\frac{1}{\sqrt{d}}\sum_{m=-J}^{J}|J,m\rangle_{1}|J,m\rangle_{2}\,.\label{eq:bipartite}\end{equation}
This state can also be written in terms of the boson operators as
\cite{multibell}\begin{equation}
|\Psi\rangle_{max}=\frac{1}{(2J)!(2J+1)^{1/2}}(\hat{a}_{1}^{\dagger}\hat{a}_{2}^{\dagger}+\hat{b}_{1}^{\dagger}\hat{b}_{2}^{\dagger})^{2J}|0\rangle\,,\label{eq:boson11}\end{equation}
where $\hat{a}_{1},\ \hat{b}_{1}$ are the two modes at site $1$,
$\hat{J}_{1}^{z}=\left(\hat{b}_{1}^{\dagger}\hat{b}_{1}-\hat{a}_{1}^{\dagger}\hat{a}_{1}\right)/2$,
with the same definition at site $2$. This equivalence is due to
the Schwinger representation, which is used to map bosonic states
into spin states. We note that due to the Schwinger representation
\cite{Sakurai}, one can write:\begin{eqnarray}
|J,m\rangle_{1} & = & \frac{\left(a_{1}^{\dagger}\right)^{J+m}\left(b_{1}^{\dagger}\right)^{J-m}}{\sqrt{\left(J+m\right)!\left(J-m\right)!}}|0_{a},0_{b}\rangle_{1}\nonumber \\
 & = & |\left(J+m\right)_{a},\left(J-m\right)_{b}\rangle_{1}\,.\label{eq:bosonic}\end{eqnarray}
Here $|\left(J+m\right)_{a},\left(J-m\right)_{b}\rangle_{1}$ represents
a general harmonic oscillator state of two bosonic or harmonic oscillator
modes localized at position $1$.} 
\item \textcolor{black}{More general, non-maximally entangled but highly
correlated spin states of form\begin{eqnarray}
|\psi\rangle_{non} & = & \frac{1}{\sqrt{n}}[r_{-J}|J,-J\rangle^{\otimes N}+r_{-J+1}|J,-J+1\rangle^{\otimes N}\nonumber \\
 &  & \ \ \ \ +...+r_{J}|J,+J\rangle^{\otimes N}]\,,\label{eq:nonmaximally state}\end{eqnarray}
where $|J,m\rangle^{\otimes N}=\Pi_{k=1}^{N}|J,m\rangle_{k}$, $n={\displaystyle \sum_{m=-J}^{J}}r_{m}^{2}$.
Here we will be restricted to the case of real parameters symmetrically
distributed around $m=0$. The amplitude $r_{m}$ can be selected
to optimize the nonlocality result. For example, with $N$ sites and
a spin-$1$ system, the state has the form\begin{eqnarray}
|\psi\rangle & = & \frac{1}{\sqrt{r^{2}+2}}(|1,-1\rangle^{\otimes N}+r|1,0\rangle^{\otimes N}\nonumber \\
 &  & \ \ \ \ +|1,+1\rangle^{\otimes N})\,,\label{eq:staterspin1}\end{eqnarray}
which has been shown by Acin et al. \cite{acinthreelevel} to give
improved violation over the maximally entangled state for some Bell
inequalities.} 
\end{enumerate}

\section{\textbf{Spin-$1/2$ case}}

\textbf{\emph{General nonlocality criteria:}} The general entanglement,
EPR steering and Bell inequalities (\ref{eq:entcrit}-\ref{eq:bellcrit})
apply in this case, in addition to the $C_{J}$-criteria (\ref{eq:spinjent})-(\ref{eq:spinjsteer})
specific to spin $J=1/2$. 

For spin-$1/2$, it is convenient to use the Pauli spin operators
$\hat{\sigma}^{x}$, $\hat{\sigma}^{y}$, $\hat{\sigma}^{z}$ so that
a connection can be made with previous criteria. The \emph{Bell inequality}
of (\ref{eq:bellcrit}) becomes\begin{equation}
|\langle\prod_{k=1}^{N}\sigma_{k}^{s_{k}}\rangle|^{2}\leq\langle\prod_{k=1}^{N}[(\sigma_{k}^{x})^{2}+(\sigma_{k}^{y})^{2}]\rangle=2^{N}\,,\label{eq:merminbell}\end{equation}
where $\hat{\sigma}_{k}^{\pm}=\hat{\sigma}_{k}^{x}+i\hat{\sigma}_{k}^{y}$.
Bell nonlocality is implied by the violation of this inequality. Similarly,
the generalised HZ \emph{entanglement} inequality of (\ref{eq:entcrit})
and (\ref{eq:entxy}) becomes \begin{eqnarray}
|\langle\prod_{k=1}^{N}\sigma_{k}^{s_{k}}\rangle|^{2} & \leq & \langle\prod_{k=1}^{N}[(\sigma_{k}^{x})^{2}+(\sigma_{k}^{y})^{2}+2l_{k}\sigma_{k}^{z}]\rangle\nonumber \\
 & = & 2^{N}\langle\prod_{k=1}^{N}[1+l_{k}\sigma_{k}^{z}]\rangle\,,\label{eq:paulientcrit}\end{eqnarray}
where $l_{k}$ and $s_{k}$ can be independently selected as $+$
or $-$, and the \emph{EPR steering} inequality (\ref{eq:sterrcrit})
becomes \begin{eqnarray}
|\langle\prod_{k=1}^{N}\sigma_{k}^{s_{k}}\rangle|^{2} & \leq & 2^{N-1}\langle(\sigma_{k}^{x})^{2}+(\sigma_{k}^{y})^{2}+2l_{k}\sigma_{k}^{z}\rangle\nonumber \\
 & = & 2^{N}\langle1+l_{k}\sigma_{k}^{z}\rangle\,.\label{eq:paulisteercrit}\end{eqnarray}

\textbf{\emph{$C_{J}$-nonlocality criteria}}: The quantum uncertainty
relation \begin{equation}
\Delta^{2}\hat{\sigma}^{x}+\Delta^{2}\hat{\sigma}^{y}\geq1\label{eq:spin1/2ur}\end{equation}
follows from $\Delta^{2}\hat{\sigma}^{x}+\Delta^{2}\hat{\sigma}^{y}+\Delta^{2}\hat{\sigma}^{z}\geq2$
\cite{hoftoth}. Hence the $C_{J}$-\emph{entanglement} inequality
of (\ref{eq:spinjent}) becomes \begin{eqnarray}
|\langle\prod_{k=1}^{N}\sigma_{k}^{s_{k}}\rangle|^{2} & \leq & \langle\prod_{k=1}^{N}[(\sigma_{k}^{x})^{2}+(\sigma_{k}^{y})^{2}-1]\rangle=1\,.\label{eq:merminent}\end{eqnarray}
The violation of this inequality thus implies entanglement. EPR-steering
is implied by violation of the \emph{EPR steering} inequality\begin{eqnarray}
|\langle\prod_{k=1}^{N}\sigma_{k}^{s_{k}}\rangle|^{2} & \leq & \langle[(\sigma_{1}^{x})^{2}+(\sigma_{1}^{y})^{2}-1]\nonumber \\
 &  & \times\prod_{k=2}^{N}[(\sigma_{k}^{x})^{2}+(\sigma_{k}^{y})^{2}]\rangle=2^{N-1}\,.\label{eq:merminsteer}\end{eqnarray}

The entanglement criterion (\ref{eq:merminent}) has been derived
by Roy \cite{royprl}, while the Bell inequality (\ref{eq:merminbell})
(when expressed in terms of the real or imaginary parts of the left
side) becomes that of Mermin \cite{mermin} (for $N$ even) and Ardehali
\cite{ard} (for $N$ odd). This Bell inequality is known to be weaker
than the full MABK Bell inequalities (that of Mermin's for $N$ odd,
and Ardehali's for $N$ even) \cite{beklys}. For EPR steering, (\ref{eq:merminsteer})
reduces to one of the EPR steering inequalities derived for the qubit
case by Cavalcanti et al. \cite{multipartitesteer}. %
\begin{figure}[b]
\begin{centering}
\includegraphics[width=0.8\columnwidth]{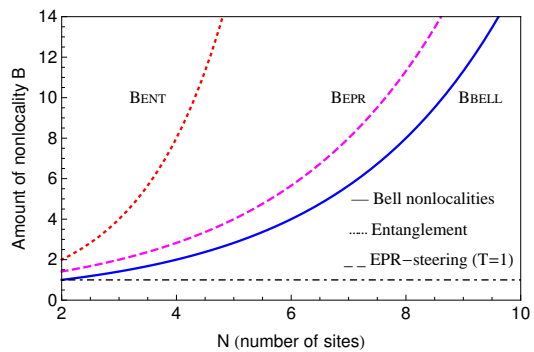} 
\par\end{centering}

\caption{(Color online) Spin $1/2$: Entanglement (dotted), EPR-steering ($T=1$)
(dashed), and Bell nonlocalities (solid) can be detected for GHZ states
of Eq. (\ref{eq:ghz}) using $C_{J}$-criteria in Eqs. (\ref{eq:merminent})
and (\ref{eq:merminsteer}) and Bell inequalities (\ref{eq:merminbell}) when
$B_{ENT},\ B_{EPR},\ B_{BELL}>1$. \textcolor{black}{$B_{ENT}$ for
the generalised HZ entanglement criterion (\ref{eq:entcrit}), i.e.,
Eq. (\ref{eq:paulientcrit}), becomes infinite in this case.}\label{fig:Spin-1/2}}

\end{figure}

\textbf{\emph{Quantum prediction for multi-site qubits:}}\textcolor{black}{
We consider the $N$-partite GHZ states, denoting $|\frac{1}{2},-\frac{1}{2}\rangle$
and $|\frac{1}{2},+\frac{1}{2}\rangle$ symbolically by $|0\rangle$
and $|1\rangle$ respectively: \begin{equation}
|\psi\rangle_{max}=\frac{1}{\sqrt{2}}(|0\rangle^{\otimes N}+|1\rangle^{\otimes N})\,.\label{eq:ghz}\end{equation}
}We define $B_{BELL},$ $B_{EPR}$, and $B_{ENT}$ for the Bell-nonlocality,
steering and entanglement inequalities respectively as the square
root of the ratio of the left side ($L$) and right side ($R$) of
the inequalities. The generalised HZ entanglement inequality (\ref{eq:entcrit}),
i.e. (\ref{eq:paulientcrit}), allows the strongest violation $B_{ENT}$
possible for this state. In fact we can consider the generalised GHZ
state \textcolor{black}{\begin{equation}
|\psi\rangle_{max}=\cos\theta|0\rangle^{\otimes N}+\sin\theta|1\rangle^{\otimes N}\,.\label{eq:ghz-2}\end{equation}
}The $l_{k}$ can be chosen so that the right side ($R$) is zero,
and the $s_{k}$ so that the left side ($L$) is non-zero. In the
spin notation, we choose specifically: \textcolor{black}{\begin{eqnarray}
L & = & |\langle\prod_{k=1}^{N}J_{k}^{-}\rangle|^{2}\nonumber \\
 & = & \Bigl|\underset{m=-J+1}{\overset{J}{\sum}}r_{m-1}^{*}r_{m}\left[(J+m)(J-m+1)\right]^{N/2}\Bigl|^{2}\nonumber \\
 & = & \left(cos\theta sin\theta\right)^{2}\,,\label{eq:lhsgenghz}\end{eqnarray}
and}\textcolor{red}{ }\textcolor{black}{\begin{eqnarray}
R & = & \langle J_{1}^{+}J_{1}^{-}\prod_{k=2}^{N}J_{k}^{-}J_{k}^{+}\rangle\nonumber \\
 & = & \sum_{m=-J}^{J}|r_{m}|^{2}(J-m)(J+m+1)\left[(J-m)(J+m+1)\right]^{N-1}\nonumber \\
 & = & 0\,,\end{eqnarray}
}where $J=1/2$, and $m=\pm1/2$. The ratio $B_{ENT}=\sqrt{L/R}\rightarrow\infty$
for choices of $\theta$ other than $0$, $\pi/2$ and the criterion
can detect all entanglement for this state. This is a stronger result
than that of \cite{mulithillery} who considered the generalised GHZ
state where the coefficients are asymmetric, but they did not consider
the generalised entanglement criterion which involves independent
choices of $l_{k}$. We note the EPR-steering inequality (\ref{eq:paulisteercrit})
reduces to the Bell inequality for these states because the correlation
$\langle\sigma_{k}^{z}\rangle=0$, and hence it is better to use the
$C_{J}$- EPR-steering criterion in this case.

The $C_{J}$-entanglement criterion (\ref{eq:spinjent}), i.e., Eq.
(\ref{eq:merminent}), can also be studied for the generalised GHZ
state, since $L$ is also given by Eq. (\ref{eq:lhsgenghz}). The
$R$ in this case (converting to the spin-$1/2$ operators), however,
is always $1/2^{2N}$, meaning that entanglement is only detected
when $\sin^{2}2\theta>2^{2}/2^{2N}=1/2^{2(N-1)}$, i.e., when $\sin2\theta>1/2^{(N-1)}$.
Considering the symmetric case for $\theta=\pi/4$, we note the $C_{J}$-criterion
is satisfied for the GHZ states for all $N\geq2$ \cite{royprl}. 

The Bell criterion is satisfied only for $N\geq3$ \cite{mermin}.
The EPR steering inequality (\ref{eq:merminsteer}) for $T=1$ is
also violated for $N\geq2$. We note the amount of violation for these
inequalities increases exponentially with $N$, as shown in Fig. \ref{fig:Spin-1/2}
$ $and as reported by Mermin \cite{mermin}, Roy \cite{royprl},
and Cavalcanti et al. \cite{multipartitesteer}. Defining $B_{T}$
to correspond to the ratio of $L$ to $R$ for the general case of
$T$ quantum sites, we note that $B_{T}=2^{(N+T-2)/2}$, so that $B_{ENT}=2^{N-1}$,
$B_{EPR}=2^{(N-1)/2}$, and $B_{BELL}=2^{(N-2)/2}$. For Bell nonlocality,
the MABK Bell inequality \cite{mermin,ard,beklys} gives stronger
violations for these GHZ states ($B_{BELL}=2^{(N-1)/2})$, and it
is violated for all $N\geq$2. The MABK inequalities involve a different
derivation, and these alternative nonlocality inequalities derived
for the qubit (spin-$1/2$) case have been studied in Ref. \cite{multipartitesteer}.

\section{\textbf{Spin-$1$ case}}

A Bell inequality for multiple spin-$1$ systems is given by Eq. (\ref{eq:bellcrit}).
The $C_{J}$-entanglement and EPR-steering criteria (\ref{eq:spinjent})
and (\ref{eq:spinjsteer}) can be used to test for the entanglement
and EPR-steering nonlocalities, which for spin-$1$ are based on the
value $C_{J}=7/16$ derived by \cite{hoftoth}. \textcolor{black}{In
this case, we find the {}``$C_{J}"$ criterion (\ref{eq:spinjent})
becomes more useful than the generalised HZ entanglement criterion
(\ref{eq:entcrit}), in the sense that the right-hand side of the
relevant inequality becomes smaller.}

\textbf{\emph{Maximally entangled state}}\textbf{\textcolor{black}{\emph{:}}}\textcolor{black}{
In this case, the inequalities can be investigated for the bipartite
case of two sites ($N=2$). Firstly the }\textcolor{black}{\emph{maximally
entangled~}}\textcolor{black}{ state (\ref{eq:bipartite}) can be
written more explicitly for spin-$1$ ($d=3$) as:\begin{align}
|\psi\rangle_{max} & =\frac{1}{\sqrt{3}}(|1,-1\rangle_{1}|1,-1\rangle_{2}+|1,0\rangle_{1}|1,0\rangle_{2}\nonumber \\
 & \ \ \ \ \ \ \ \ \ +|1,+1\rangle_{1}|1,+1\rangle_{2})\,.\label{eq:qudit st 1}\end{align}
This state can also be written in terms of the boson operators as
\cite{multibell}\begin{eqnarray}
|\Psi\rangle_{max} & = & \frac{1}{2\sqrt{3}}(a_{1}^{\dagger}a_{2}^{\dagger}+b_{1}^{\dagger}b_{2}^{\dagger})^{2}|0\rangle\,.\end{eqnarray}
No violation of the Bell inequality (\ref{eq:bellcrit}) is observed
here, but the $C_{J}$-entanglement criterion (\ref{eq:spinjent})
is satisfied.}

\begin{figure}[t]
\begin{centering}
\includegraphics[width=0.8\columnwidth]{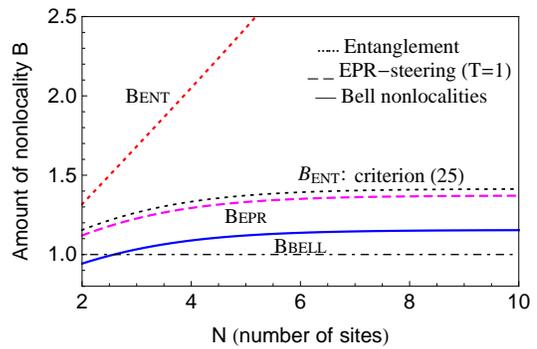}
\par\end{centering}

\centering{}\caption{(Color online) Spin-$1$ case for the maximally entangled state (\ref{eq:maximally entangled state}).
Nonlocality detected by the Bell inequality (\ref{eq:bellcrit}) and
the $C_{J}$ entanglement and EPR-steering criteria (\ref{eq:spinjent})
and (\ref{eq:spinjsteer}). The nonlocality is detected when the appropriate
$B>1$. \label{fig:Spin-1}\textcolor{black}{ The entanglement measured
by the generalised HZ entanglement criterion (\ref{eq:entcrit}) is
plotted for comparison.}}

\end{figure}

\textcolor{black}{Allowing for more sites ($N>2$), we consider the
state (\ref{eq:maximally entangled state}) that can be written equivalently
as:}\begin{equation}
|\Psi\rangle_{max}=norm(a_{1}^{\dagger}a_{2}^{\dagger}...a_{N}^{\dagger}+b_{1}^{\dagger}b_{2}^{\dagger}...b_{N}^{\dagger})^{2J}|0\rangle\label{eq:bosonmultisite}\end{equation}
with $J=1$. Violation of the Bell's inequality (\ref{eq:bellcrit})
is possible for this state with $N>2$, as shown in Fig. \ref{fig:Spin-1}.
The limit of the square root of the ratio of the left side ($L$)
to the right side ($R$) of the Bell criterion (\ref{eq:bellcrit})
as $N\rightarrow\infty$ is $2/\sqrt{3}$. In fact generally the ratio
is given as:\begin{equation}
B_{BELL}=\frac{2\left(2^{N-1}\right)^{1/2}}{\sqrt{3}\left(2^{N-1}+1\right)^{1/2}}\,.\label{eq:ratioBbellspin1}\end{equation}
This ratio may be compared with the $C_{J}$-criterion (\ref{eq:spinjent})
for entanglement:\begin{equation}
B_{ENT}=\frac{2^{3N}}{\left(3^{2N}2^{N-1}+5^{2N}\right)^{1/2}\left(2^{N-1}+1\right)^{1/2}}\,.\label{eq:ratiobentspin1}\end{equation}
Entanglement can be proved for all $N\geq2$. The ratio $B_{ENT}$
increases with $N$, and is favourable compared to that obtained from
generalised HZ entanglement criterion (\ref{eq:entcrit}) (Fig. \ref{fig:Spin-1}).
The amount of nonlocality for the asymmetric EPR-steering case ($T=1$)
is also plotted. EPR-steering can be detected via the $C_{J}$-criterion
(\ref{eq:spinjsteer}) when $B_{EPR}=2^{(N+4)/2}/\left[\sqrt{17}(2^{N-1}+1)^{1/2}\right]>1$
for $T=1,\ N\geq2$. 

\begin{figure}[t]
\begin{centering}
\includegraphics[width=0.8\columnwidth]{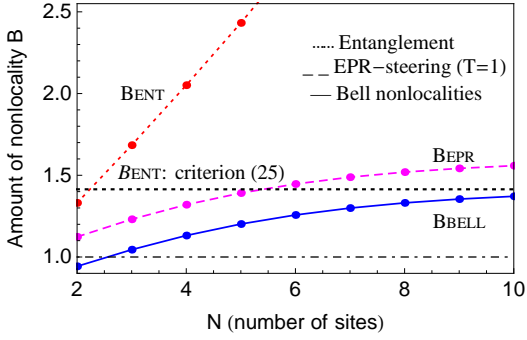} 
\par\end{centering}

\caption{(Color online) Spin-$1$ case for the non-maximally entangled state
(\ref{eq:staterspin1}) with optimal $r.$ The amount of nonlocality
$B_{BELL}$ (solid), $B_{EPR}\ (T=1)$ (dashed), \textcolor{black}{and
$B_{ENT}$ (dotted) detected by the Bell inequality (\ref{eq:bellcrit})
and $C_{J}$-criteria (\ref{eq:spinjsteer}) and (\ref{eq:spinjent})
is larger than for state (\ref{eq:maximally entangled state}) shown
in Fig. \ref{fig:Spin-1}. Here, $B_{BELL},\ B_{EPR},\ B_{ENT}>1$
implies the detection of the respective nonlocality. The entanglement
detected by the generalised HZ entanglement criterion (\ref{eq:entcrit})
with optimal $r$ is shown in here for comparison. \label{fig:nonmaximally entangled state spin-1}}}

\end{figure}

\textbf{\emph{Non-maximally entangled state: }}We can also consider
the more general state of Eq. (\ref{eq:staterspin1}). In this case,
the left ($L$) and right ($R$) sides of the Bell inequality (\ref{eq:bellcrit})
become:\begin{align}
L & =\frac{2^{N+2}r^{2}}{(r^{2}+2)^{2}}\,,\label{eq:ratioL}\\
R & =\frac{2^{N}r^{2}+2}{r^{2}+2}\,,\label{eq:ratior1}\end{align}
then\begin{equation}
B_{BELL}=\frac{2^{(N+2)/2}r}{(r^{2}+2)^{1/2}\left[2^{N}r^{2}+2\right]^{1/2}}\,.\label{eq:ratio}\end{equation}
Optimising $r$ for each value of $N$, we can get a violation from
$N=3$ sites for spin $1$, as for the maximally entangled state,
but the violations are greater. The amount of violation is $B_{BELL}\rightarrow\sqrt{2}$
as the number of sites $N$ increases (Fig. \ref{fig:nonmaximally entangled state spin-1}).

The result for the $C_{J}$-entanglement inequality (\ref{eq:spinjent})
is\begin{equation}
B_{ENT}=\frac{2^{(N+2)/2}r}{(r^{2}+2)^{1/2}\left[\left(25/16\right)^{N}r^{2}+2\left(9/16\right)^{N}\right]^{1/2}}\,,\end{equation}
which increases with number of sites $N$. The results are shown in
Fig. \ref{fig:nonmaximally entangled state spin-1}.

\textcolor{black}{The value of $B$ for the $C_{J}$-criterion for
EPR-steering (\ref{eq:spinjsteer}) can be derived as\begin{equation}
B_{EPR}=\frac{2^{(N+2)/2}r}{(r^{2}+2)^{1/2}\left[\left(9/8\right)^{N}+r^{2}2^{N-5}25\right]^{1/2}}\,,\end{equation}
as shown in Fig. }\ref{fig:nonmaximally entangled state spin-1}\textcolor{black}{.
More generally, using $C_{J}$-criterion (\ref{eq:spinjsteer-1}),
the result is\begin{equation}
B_{T}=\frac{2^{(N+2)/2}r}{(r^{2}+2)^{1/2}\left[2\left(9/16\right)^{T}+r^{2}2^{N-T}\left(25/16\right)^{T}\right]^{1/2}}\,.\end{equation}
}

\section{Spin-$J$ Case}

\subsection{Bell nonlocality}

The maximally entangled state (\ref{eq:bosonmultisite}) gives violation
of the Bell inequalities (\ref{eq:bellcrit})$ $ only \textcolor{black}{if
$d=2,\ 3$ and $N\geq3$.} However, the Bell inequalities can be violated
for larger $d$ by the optimally selected symmetric non-maximally
entangled states (\ref{eq:nonmaximally state}), provided the number
of sites $N$ is high enough. Figure 6 shows the results for $d=2,...,7$.
In this section we treat cases of arbitrary spin-$J$.

\textcolor{blue}{}%
\begin{figure}[t]
\begin{centering}
\includegraphics[width=0.8\columnwidth]{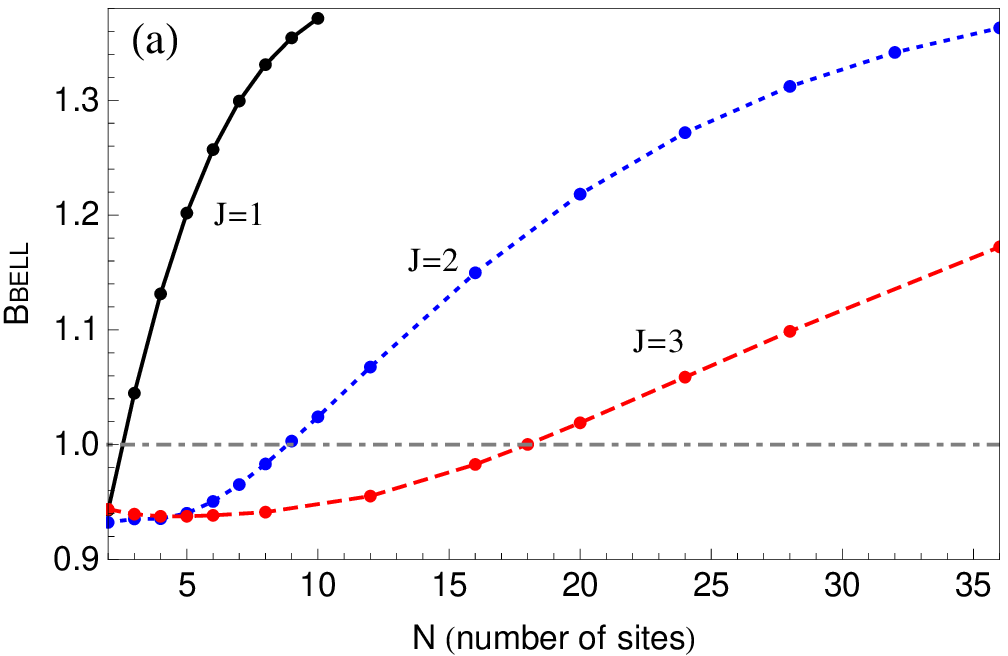}
\par\end{centering}

\begin{centering}
\includegraphics[width=0.8\columnwidth]{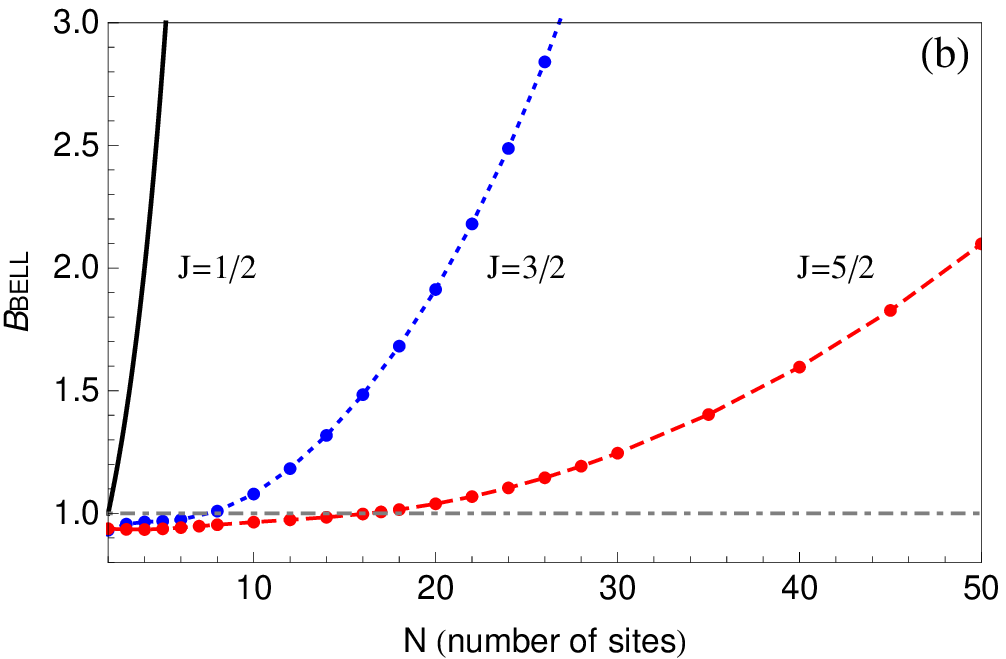}
\par\end{centering}

\textcolor{black}{\caption{(Color online)\textcolor{black}{ Violation of the Bell inequality
(\ref{eq:bellcrit}) as measured by $B_{BELL}$ versus $N$ for the
symmetric non-maximally entangled state (\ref{eq:nonmaximally state}):
(a) $J=1,\ 2,\ 3$ and (b) $J=1/2,\ 3/2,\ 5/2$. Bell nonlocality
is confirmed when $B_{BELL}>1$.}\textcolor{blue}{\label{fig:growth_ratio_ent-1}}}
}
\end{figure}

\begin{figure}[t]
\begin{centering}
\includegraphics[width=0.8\columnwidth]{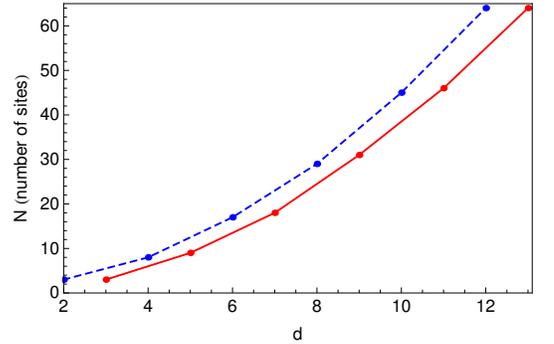}
\par\end{centering}

\caption{\textcolor{black}{(Color online) Minimum number of sites $N$ needed
for a given $d$ ($d=2J+1$) in order to violate the Bell inequality
(\ref{eq:bellcrit}) for a non-maximally entangled state (\ref{eq:nonmaximally state}):
even $d$ (dashed curve), odd $d$ (solid curve).}\textcolor{red}{
}\textcolor{black}{\label{fig:minimum N for given d}}}

\end{figure}

For non-maximally entangled states (\ref{eq:nonmaximally state}),
first we need to calculate the values:\begin{align}
\langle(J_{1}^{z})^{2}\rangle & =\frac{1}{n}\sum_{m=-J}^{J}m^{2}r_{m}^{2}\,,\nonumber \\
\langle(J_{1}^{z})^{2}(J_{2}^{z})^{2}\rangle & =\frac{1}{n}\sum_{m=-J}^{J}m^{4}r_{m}^{2}\,,\end{align}
where $n={\displaystyle \sum_{m=-J}^{J}}r_{m}^{2}$, $m=-J,...,+J$.
Then we obtain:\textcolor{black}{ \begin{eqnarray}
L & = & |\langle\prod_{k=1}^{N}J_{k}^{-}\rangle|^{2}\nonumber \\
 & = & \frac{1}{n^{2}}\left[\underset{m=-J}{\overset{J}{\sum}}r_{m}r_{m+1}(\sqrt{J-m}\sqrt{J+m+1})^{N}\right]^{2}\,,\\
R & = & \langle\left[\left(J_{1}\right)^{2}-\left(J_{1}^{z}\right)^{2}\right]...\left[\left(J_{N}\right)^{2}-\left(J_{N}^{z}\right)^{2}\right]\rangle\nonumber \\
 & = & \frac{1}{n}\underset{m=-J}{\overset{J}{\sum}}r_{m}^{2}\left[J(J+1)-m^{2}\right]^{N}\,,\end{eqnarray}
\begin{equation}
B_{BELL}=\frac{\underset{m=-J}{\overset{J}{\sum}}r_{m}r_{m+1}(\sqrt{J-m}\sqrt{J+m+1})^{N}}{\left\{ n\underset{m=-J}{\overset{J}{\sum}}r_{m}^{2}\left[J(J+1)-m^{2}\right]^{N}\right\} {}^{1/2}}\,.\label{eq:bellspinj}\end{equation}
}Optimizing the value of $r_{m}$ for each spin value $J$ and number
of sites $N$, we see that the Bell inequalities can be violated ($B_{BELL}>1$)
for larger $d$ provided $N$ is sufficiently large (Fig. \ref{fig:minimum N for given d}).

\textcolor{blue}{}%
\begin{figure}[t]
\begin{centering}
\includegraphics[width=0.8\columnwidth]{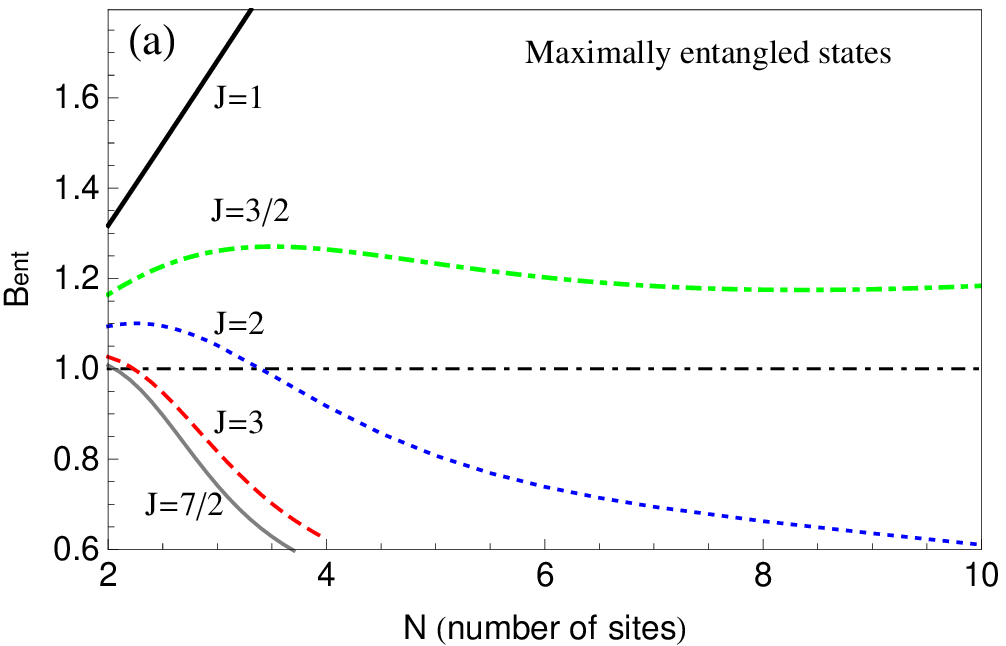}
\par\end{centering}

\begin{centering}
\includegraphics[width=0.8\columnwidth]{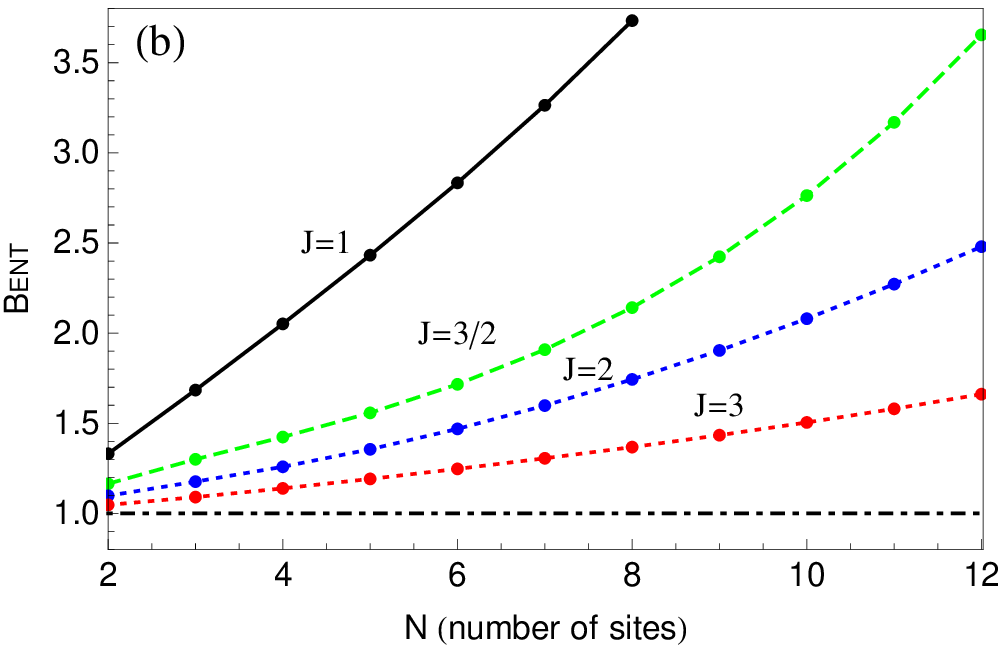}
\par\end{centering}

\textcolor{black}{\caption{(Color online)\textcolor{black}{ Entanglement as measured by the $C_{J}$-criterion
(\ref{eq:spinjent}): $B_{ENT}$ versus $N$ for (a) the maximally
entangled state (\ref{eq:maximally entangled state}) and (b) the
optimal non-maximally entangled state (\ref{eq:nonmaximally state}).
Entanglement is confirmed when $B_{ENT}>1$.}\textcolor{blue}{\label{fig:growth_ratio_ent}}}
}
\end{figure}

\subsection{Entanglement}

\textcolor{black}{The left side ($L$) of the inequality for the generalised
HZ entanglement criterion (\ref{eq:entcrit}) and the $C_{J}$-entanglement
criterion (\ref{eq:spinjent}) is the same as that for the Bell inequality;
the right side ($R$), however, changes. Entanglement can be proved
for the state (\ref{eq:nonmaximally state}) via the $C_{J}$- entanglement
criterion (\ref{eq:spinjent}) for spin-$J$ when\begin{equation}
B_{ENT}=\frac{\underset{m=-J}{\overset{J}{\sum}}r_{m}r_{m+1}(\sqrt{J-m}\sqrt{J+m+1})^{N}}{[n\underset{m=-J}{\overset{J}{\sum}}r_{m}^{2}\left[J(J+1)-m^{2}-C_{J}\right]^{N}]^{1/2}}>1\,.\end{equation}
For maximally entangled states, $r_{m}$ is fixed as $r_{m}=\left[\sqrt{\left(J-m\right)!(J+m)!}\right]^{N-2}$,
and then the criterion is only satisfied for lower $J<4$ and increases
or is steady with $N$ only for $J<2$ (Fig. \ref{fig:growth_ratio_ent}(a)).
However, for the symmetric non-maximally entangled states (\ref{eq:nonmaximally state}),
$B_{ENT}>1$ can occur for all spin $J$ and $N$, provided the amplitudes
$r_{m}$ are optimally chosen. For fixed $J$, the value of $B_{ENT}$
increases with $N$, while for fixed $N$ the violation decreases
with increasing $J$ (Fig. \ref{fig:growth_ratio_ent}(b)). }

\begin{figure}[t]
\begin{centering}
\includegraphics[width=0.8\columnwidth]{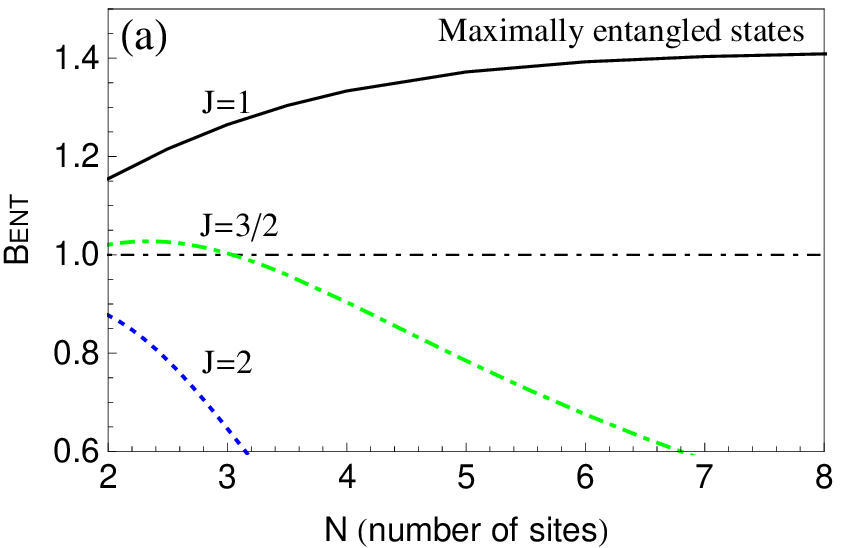}
\par\end{centering}

\begin{centering}
\includegraphics[width=0.8\columnwidth]{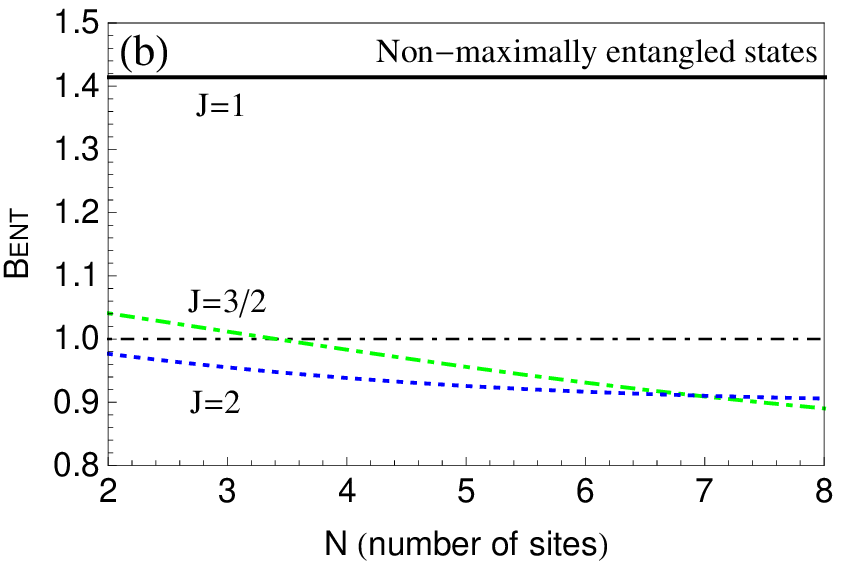}
\par\end{centering}

\textcolor{black}{\caption{\textcolor{black}{(Color online) Entanglement as measured by the generalised
HZ entanglement criterion (\ref{eq:entcrit}): $B_{ENT}$ versus $N$
for (a) the maximally entangled state (\ref{eq:maximally entangled state}),
and (b) the optimal non-maximally entangled state (\ref{eq:nonmaximally state}).
Entanglement is confirmed when $B_{ENT}>1$.}\textcolor{blue}{\label{fig:figure for criterion(25)}}}
}
\end{figure}

\textcolor{black}{The generalised HZ entanglement criterion (\ref{eq:entcrit}),
however, has a different $R$ (we choose appropriate $s_{k},\ l_{k}$
to get larger $L$ and smaller $R$): \begin{align}
L & =|\langle\prod_{k=1}^{N}J_{k}^{-}\rangle|^{2}\nonumber \\
 & =\Bigl|\underset{m=-J+1}{\overset{J}{\sum}}\frac{r_{m-1}r_{m}}{n}\left[(J+m)(J-m+1)\right]^{N/2}\Bigl|^{2}\,,\end{align}
and\begin{align}
R & =\langle J_{1}^{+}J_{1}^{-}\prod_{k=2}^{N}J_{k}^{-}J_{k}^{+}\rangle\nonumber \\
 & =\sum_{m=-J}^{J}\frac{r_{m}^{2}}{n}\times\nonumber \\
 & \{(J-m)(J+m+1)\left[(J-m)(J+m+1)\right]^{N-1}\}\,,\end{align}
so that entanglement is detected when this $B_{ENT}=\sqrt{L/R}>1$.
For the maximally entangled state, entanglement is only detected for
lower $J$, as shown in Fig. \ref{fig:figure for criterion(25)} (a),
while the result for symmetric but otherwise optimised $r_{m}$ is
shown in Fig. \ref{fig:figure for criterion(25)} (b). The generalised
HZ entanglement criterion becomes less effective than the $C_{J}$-entanglement
criterion for $J>1/2$.}

\textcolor{blue}{}%

\textbf{\textcolor{black}{\emph{Entanglement for the bipartite spin-$J$
case: }}}\textcolor{black}{The bipartite ($N=2$) case for arbitrary
spin $J$ has been considered recently by Zheng et al. \cite{spinhillery}
using criteria similar to (\ref{eq:entcrit}) (but with restricted
choices of $l_{k}$). Here we use the generalised HZ entanglement
criterion (\ref{eq:entcrit}): entanglement is detected when \begin{equation}
|\langle J_{1}^{s_{1}}J_{2}^{s_{2}}\rangle|^{2}>\langle J_{1}^{l_{1}}J_{1}^{-l_{1}}J_{2}^{l_{2}}J_{2}^{-l_{2}}\rangle\,,\label{eq:zdhbipartitespinJ}\end{equation}
where $s_{k}$ and $l_{k}$ are independently chosen to be $+$ or
$-$. We consider the highly correlated state (\ref{eq:nonmaximally state})
\begin{equation}
|\psi\rangle=\frac{1}{n}\sum_{m=-J}^{J}r_{m}|J,m\rangle_{1}|J,m\rangle_{2}\label{eq:bipartite maximally}\end{equation}
for which \begin{equation}
L=|\langle J_{1}^{-}J_{2}^{-}\rangle|^{2}=\frac{1}{n^{2}}\Bigl|\underset{m=-J+1}{\overset{J}{\sum}}r_{m-1}r_{m}(J+m)(J-m+1)\Bigl|^{2}\,,\end{equation}
and\begin{eqnarray}
R & = & \langle J_{1}^{+}J_{1}^{-}J_{2}^{-}J_{2}^{+}\rangle\nonumber \\
 & = & \frac{1}{n}\sum_{m=-J}^{J}r_{m}^{2}(J^{2}-m^{2})[(J+1)^{2}-m^{2}]\,.\end{eqnarray}
Immediately, for $J=1/2$, we see that $R$ is zero for all choices
of $r_{m}$, i.e. for the generalised Bell state\begin{equation}
|\psi\rangle_{max}=\cos\theta|0\rangle^{\otimes2}+\sin\theta|1\rangle^{\otimes2})\,,\label{eq:ghz-1}\end{equation}
and hence the criterion detects }\textcolor{black}{\emph{all }}\textcolor{black}{entanglement
for this state. This contrasts with the criterion considered by HZ
\cite{mulithillery}, which does not detect entanglement for the symmetric
case $\cos\theta=\sin\theta$. }

\textcolor{black}{For spin $J=1$, detection is still possible though
less ideal, as shown for the choice of constant and real $r_{m}$
(a maximally entangled state) in Fig. \ref{fig:Spin-1} and for an
optimally chosen but real and symmetric $r_{m}$ in Fig. \ref{fig:nonmaximally entangled state spin-1}. }

\begin{figure}
\begin{centering}
\includegraphics[width=0.8\columnwidth]{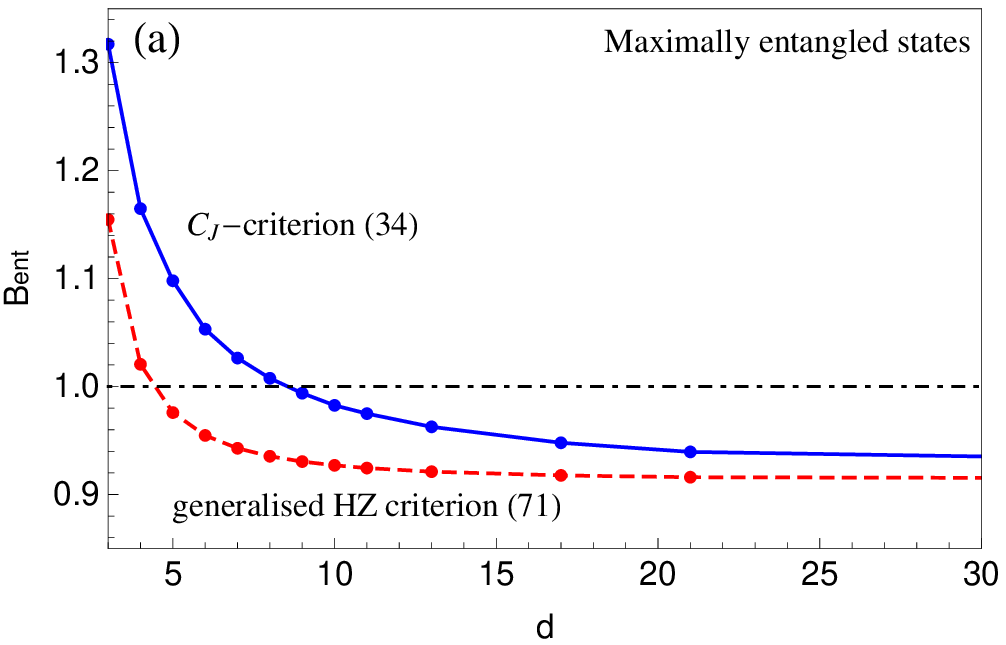}
\par\end{centering}

\begin{centering}
\includegraphics[width=0.8\columnwidth]{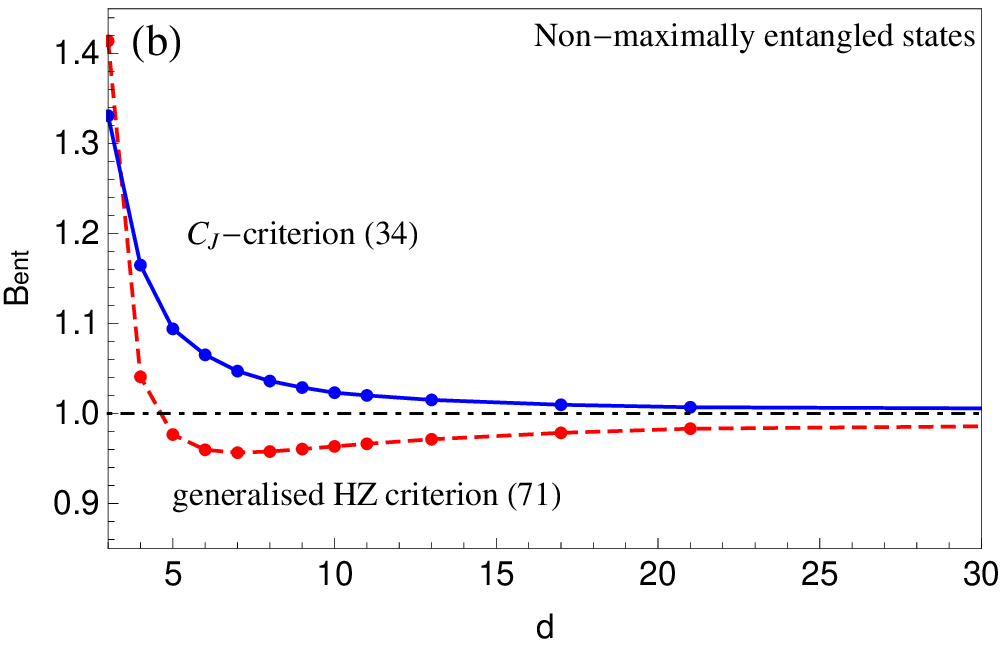}
\par\end{centering}

\caption{\textcolor{black}{(Color online) Bipartite case: Entanglement as measured
by the generalised HZ criterion (\ref{eq:zdhbipartitespinJ}) (dashed
curve) and the $C_{J}$-criterion (solid curve): $B_{ENT}$ versus
$d=2J+1$ for (a) the bipartite maximally entangled state (\ref{eq:maximally entangled state})
and (b) the non-maximally entangled state (\ref{eq:nonmaximally state})
with optimal $r_{m}$. Entanglement is confirmed when $B_{ENT}>1$.\label{fig:bipartite spin J}}}

\end{figure}

\textcolor{black}{Figure \ref{fig:bipartite spin J} presents results
for detection of entanglement for the bipartite case with increasing
$J$ for the generalised HZ entanglement criterion (\ref{eq:zdhbipartitespinJ})
and the $C_{J}$-criterion (\ref{eq:spinjent}). Neither criterion
can detect entanglement of the maximally entangled state (\ref{eq:maximally entangled state})
for high $J$. The $C_{J}$-criterion can be used to detect entanglement
for all $J$ that we have calculated using the optimised states (\ref{eq:nonmaximally state}).
These criteria may be compared with the variance Local Uncertainty
Relation (LUR) criteria of \cite{hoftoth,cj} which detect entanglement
for all the highly correlated states (\ref{eq:maximally entangled state})
and (\ref{eq:nonmaximally state}) of arbitrary $J$. We note the
earlier spin squeezing criteria of \cite{Anders_entanglement} are
not sensitive to entanglement in cases where $\langle J_{Z}\rangle=0$.}

\subsection{\textcolor{black}{EPR steering}}

\textcolor{black}{For the non-maximally entangled state (\ref{eq:nonmaximally state}),
the $C_{J}$-criterion for EPR-steering (\ref{eq:spinjsteer}) is
satisfied when\begin{align}
 & B_{EPR}=\nonumber \\
 & \frac{\underset{m=-J}{\overset{J}{\sum}}r_{m}r_{m+1}(\sqrt{J-m}\sqrt{J+m+1})^{N}}{n^{1/2}\underset{m=-J}{\overset{J}{\sum}}r_{m}^{2}\left\{ [J(J+1)-m^{2}]^{N-1}\left[J(J+1)-m^{2}-C_{J}\right]\right\} ^{1/2}}\nonumber \\
 & >1\,.\end{align}
This is predicted for all dimensions $d$ with optimal $r_{m}$, provided
the number of sites $N$ is high enough. For the more general model
(\ref{eq:sepent-1}) of $T$ quantum sites, the $C_{J}$-criterion
(\ref{eq:spinjsteer-1}) for the nonlocality becomes, for spin-$J$:\begin{align}
 & B_{T}=\nonumber \\
 & \frac{\underset{m=-J}{\overset{J}{\sum}}r_{m}r_{m+1}(\sqrt{J-m}\sqrt{J+m+1})^{N}}{[n\underset{m=-J}{\overset{J}{\sum}}r_{m}^{2}\left\{ [J(J+1)-m^{2}]^{N-T}\left[J(J+1)-m^{2}-C_{J}\right]^{T}\right\} ]^{1/2}}\nonumber \\
 & >1\,.\label{eq:cjT}\end{align}
This more general criterion is also verified for all dimensions $d$
with optimal $r_{m}$, provided the number of sites $N$ is high enough.}

\begin{figure}[t]
\begin{centering}
\includegraphics[width=0.8\columnwidth]{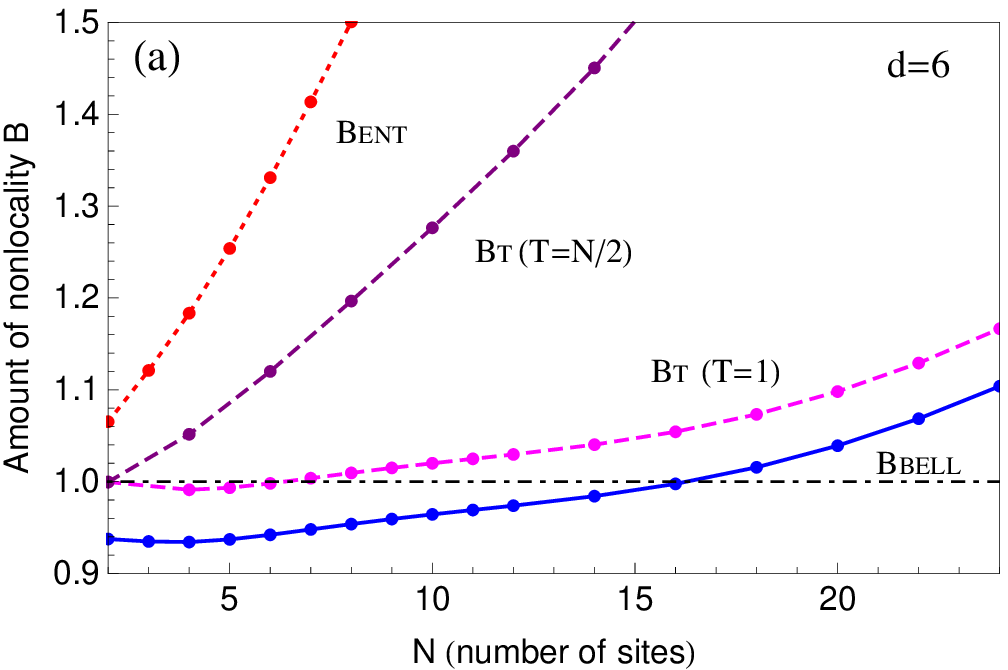}
\par\end{centering}

\begin{centering}
\includegraphics[width=0.8\columnwidth]{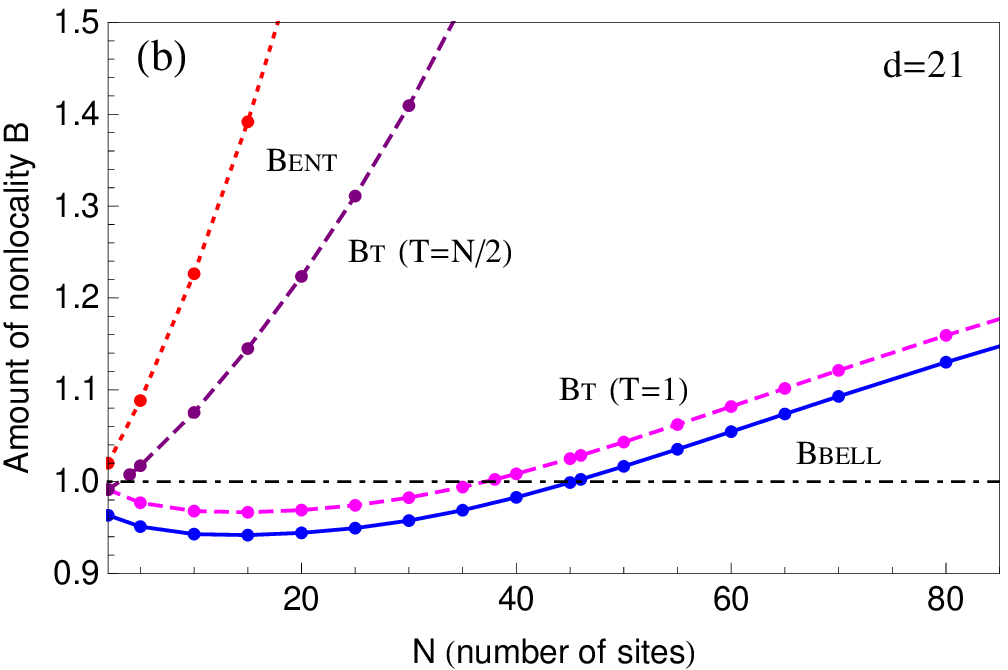}
\par\end{centering}

\caption{\textcolor{black}{(Color online) Effect of increasing the number of
sites $N$ for fixed $d$ on the strength $B$ of the nonlocality
as measured by the Bell inequality (\ref{eq:bellcrit}) and the $C_{J}$-criteria
(\ref{eq:spinjent}), (\ref{eq:spinjsteer}), and (\ref{eq:spinjsteer-1}):
(a) $d=6$ ($J=5/2$) and (b) $d=21$ ($J=10$). Results are for the
non-maximally entangled states (\ref{eq:nonmaximally state}): entanglement
(dotted) can be detected when $B_{ENT}>1$; Bell nonlocalities (solid)
can be detected when $B_{BELL}>1$. The $C_{J}$-criterion (\ref{eq:spinjsteer-1})
for the more general model of $T$ quantum sites can be verified when
$B_{T}>1$. \label{fig:Fixed-d}}}

\end{figure}
\textcolor{black}{ }%
\begin{figure}[t]
\begin{centering}
\includegraphics[width=0.8\columnwidth]{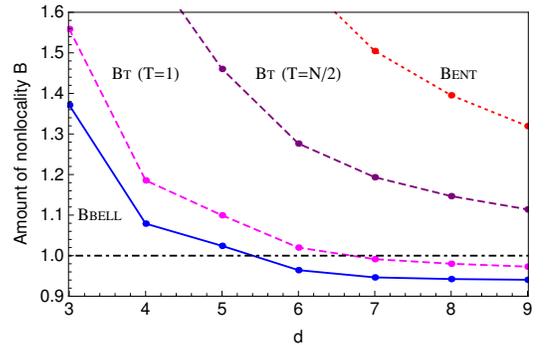}
\par\end{centering}

\caption{(Color online) The entanglement, EPR steering and Bell nonlocalities
versus $d$ (i.e., $d=2J+1$) as measured by the $C_{J}$-criteria
(\ref{eq:spinjent}), (\ref{eq:spinjsteer}) and the Bell inequality
(\ref{eq:bellcrit}) for the non-maximally entangled states (\ref{eq:nonmaximally state}):
$N=10$. The nonlocality is confirmed when $B>1$. \label{fig:Fixed-N}}

\end{figure}

\subsection{\textcolor{black}{Summary}\textcolor{black}{\emph{ }}}

\textcolor{black}{Plots of the violation of the relevant inequalities
for the three types of nonlocality for fixed $d$ and increasing $N$
are shown in Fig. \ref{fig:Fixed-d}. The strength of the violation
as measured by $B$ for these particular inequalities increases with
$N$, but this occurs for all $d$ only for the optimised non-maximally
entangled states. This effect is similar to that reported for the
MABK-type Bell inequalities of Cabello \cite{Cabello}, though an
increase in violation with $N$ was not reported for the multipartite
qudit Bell inequalities of Chen and Deng \cite{multisitequdit}. }

\textcolor{black}{For fixed $N$ and increasing $d$ ($J$), the strength
of the violation reduces (Fig. \ref{fig:Fixed-N}). This result differs
from that of Collins et al. \cite{qudit} for $N=2$, who obtained
steady violation for increasing $d$ for the maximally entangled states.
Cabello \cite{Cabello} also reported a steady violation with increased
$d$ with any fixed $N$, but this effect was not observed by the
violations of the inequalities of Son et al. \cite{multisitequditson}. }

\section{\textcolor{black}{Conclusion}}

\textcolor{black}{We have derived a unified set of measurement-based
criteria for multipartite entanglement, steering, and Bell nonlocality
for $N$-site systems of higher dimensionality $d$. Direct application
of the Bell inequality (\ref{eq:bellcrit}) shows that demonstrations
of Bell nonlocality are possible for maximally entangled highly correlated
states $|\psi_{m}\rangle$ where $d=2,\ 3$, and all $N\geq3$. Symmetric
non-maximally entangled but highly correlated states of the type considered
by Acin et al. \cite{acinthreelevel} show violations for all higher
$d$, provided the number of sites $N$ is large enough and the state
is optimised.}

\textcolor{black}{Our work also includes the derivation of entanglement
and steering criteria that take a very similar form to the Bell inequality.
We have introduced two types of such criteria for entanglement. One
is valid for all spin states, and for entanglement is similar to criteria
that have been presented by Hillery and co-workers \cite{hillzub,hilDN,mulithillery}.
We therefore call this a generalised HZ criterion. The other (a {}``$C_{J}$-criterion'')
is valid for states with a fixed total spin $J$, and reduces to the
entanglement criterion of Roy \cite{royprl} for $J=1/2$. For maximally
entangled states $|\psi_{m}\rangle$ the $C_{J}$-entanglement criterion
can only detect entanglement for low spin $J$ ($J\leq3)$. The violation
of the $C_{J}$-inequality increases with $N$ for $J=1/2$, and $J=1$,
$\ $but for higher-spin it decreases with $N$, so that the bipartite
case is optimal. The generalised HZ entanglement criterion, in a form
different from that considered originally in \cite{mulithillery},
is remarkably sensitive in the spin-$1/2$ case, being able to detect
all entanglement of a generalised multipartite GHZ state. This criterion,
however, becomes generally less sensitive than the $C_{J}$-criterion
for higher $J$. Both entanglement criteria can detect entanglement
for some optimised symmetric non-maximally entangled states, but the
first criterion is only sensitive for low $J$. Violation of the appropriate
$C_{J}$-entanglement inequality is possible in this case for all
$J$ and $N$: the violation decreases for increasing $J$, but will
increase with $N$ for fixed $J$.}

The degree of violation obtained from these Bell inequalities shows
the MABK-type growth of violation with $N$, but the violation decreases
with increasing $J$. However, for fixed $J$, one can achieve a violation
by increasing $N$ sufficiently. Our approach has the advantage that
it readily gives entanglement and steering-EPR paradox criteria and
gives analytical predictions for simple quantum states. It might be
noted that the different form of the right-hand side of the Bell and
nonlocality CFRD-type inequalities of this paper may mean a more advantageous
result for other scenarios, such as where loss is included, as studied
in \cite{function cfrd}.

As a last point, while the nonlocality criteria of this paper refer
to multipartite scenarios, they do not necessarily detect a genuine
multipartite entanglement or Bell's nonlocality, that is shared between
all $N$ parties. Such entanglement is crucial in addressing the real
existence of macroscopic entanglement, and will be treated elsewhere.

We are grateful for support from the Australian Research Council via
ACQAO COE and Discovery grants. One of us (P.D.D.) thanks the Humboldt
Society for financial support.

\end{document}